\documentclass[aps,twocolumn,showpacs,amsmath,amsfonts,amssymb]{revtex4-2}
\usepackage{graphicx}
\usepackage{epsf}
\usepackage{epstopdf}
\usepackage{xcolor}
\begin{document}

\title{Engineering Entangled Schr{\"o}dinger Cat States of Separated Cavity Modes in Cavity-QED}
\author{Abdul Q. Batin$^{\mathrm{1}}$}
\author{Suranjana Ghosh$^{\mathrm{2}}$}
\author{Utpal Roy$^{\mathrm{1}}$}
\email{uroy@iitp.ac.in}
\author{David Vitali$^{\mathrm{3}}$}
\email{david.vitali@unicam.it}

\affiliation{$^{\mathrm{1}}$Indian Institute of Technology Patna,
Bihta, Patna-801103, India\\$^{\mathrm{2}}$Indian Institute of
Science Education and Research Kolkata, Mohanpur-741246, India\\$^{\mathrm{3}}$School of Science and Technology, Physics Division, University of Camerino, Camerino, Italy}

\begin{abstract}

We provide a scheme by utilizing a two-cavity setup to generate useful quantum mechanically entangled states of two cavity fields, which themselves are prepared in Schr\"{o}dinger cat states. The underlying atom-field interaction is considered off-resonant and three atoms are successively sent through the cavities, initially fed with coherent fields. Analytical solution of the protocol, followed by conditional measurements on the atoms, produce a family of eight such entangled states. Entanglement properties of the obtained states are characterized by the Von Neumann entropy. We reveal the parameter domain for tuning the entanglement, the prime tuning parameters being the atom-field interaction time and the field amplitudes. The parameter domains for both quasi-Bell and non quasi-Bell states are discussed. We also present a Wigner phase space representation of the reduced state of the cavity, showing negative values and interference patterns similar to those of a compass state, used in quantum precision measurements, and despite its large entropy.

\end{abstract}

\maketitle

\section{Introduction}
Two noteworthy concepts fundamentally distinguish quantum mechanics from classical physics: the ideas of quantum superposition and quantum entanglement. Distinct quantum states can coexist until being measured and they can be also made nonseparable with correlated measurement outcomes
Einstein negated the idea of non-local interaction which was initially thought to occur faster than the speed of light and it was coined as ``spooky action at a distance" \cite {einstein1935can}. Later, successive experiments have been performed to demonstrate that quantum entanglement is indeed a genuine occurrence \cite{clauser1978bell,aspect1982experimental,mair2001entanglement,greenberger1990bell,erhard2020advances}. Combination of quantum entanglement and quantum superposition has become an essential tool for various emerging applications in quantum technology. There have been continuous searches for appropriate quantum superposition states, many of which are found very useful, such as Schr{\"o}dinger cat state, compass state, benzene-like state, superposition of two mode coherent states, quantum tetrachotomous state, hypercube state, superposition of single or multiple photon added coherent states \emph{etc} \cite{shukla2019quantum,roy2009sub}. Their 
non-classical features have been intensively studied \cite{buvzek1992superpositions,sanders2012review,sun1992generation,
yurke1986generating,agarwal1992new,
solomon1994characteristic,wigner1997quantum,
hillery1984distribution,kenfack2004negativity}, and they have been shown to be beneficial for a wide range of quantum technology applications, such as quantum teleportation \cite{van2001entangled,horoshko2019quantum}, quantum sensing \cite{ghosh2006mesoscopic,ghosh2009sub,roy2009sub,
ghosh2014enhanced,ghosh2019sub,akhtar2021sub,
agarwal2022quantifying,bera2020matter,bera2022quantum,
joo2011quantum,cappellaro2005entanglement,demkowicz2014using,
huang2016usefulness}, quantum communication \cite{cleve1997substituting,bostrom2002deterministic,
hastings2009superadditivity}, quantum cryptography \cite{Otta,tittel2000quantum,yin2020entanglement,
jennewein2000quantum}, and quantum computing \cite{zidan2020novel,li2014triple,ding2007review,ottaviani2010,mirrahimi2023}. Highly nontrivial, but interesting properties are displayed by various entangled states, which include the two-qubit EPR states \cite{lo2001concentrating}, the three-qubit GHZ states \cite{dur2000three}, W-states \cite{acin2001classification}, the four-qubit cluster-type states \cite{nielsen2006cluster,
walther2005experimental,briegel2001persistent}, each with their unique properties and manifestations.

The investigation and manipulation of the interaction between light and matter in systems, such as cavity QED, NMR, molecules, and ultracold atoms provide favourable platforms to perform tasks in quantum technology and quantum logic operations. The strong coupling regime, achievable in cavity QED, allows for the exploration of the light and matter interaction in a confined space \cite{haroche1989cavity,wang2019turning}, making it an intriguing area of study in quantum optics. Particularly, Haroche and co-workers \cite{davidovich1996mesoscopic,brune1996observing,
raimond1997reversible,raimond2001manipulating} have demonstrated the production of mesoscopic superposition states in high quality cavity QED setup with large dispersive or resonant interaction, which is exploited to generate various non-classical superposition and entanglement of coherent states \cite{agarwal1997atomic,gerry1996generation,
montina2002bistability,solano2003generalized}.

In this work, we exploit the above mentioned two-cavity setup to provide a quantum circuit, comprising of two microwave cavities, Ramsey zones, atomic sources and detectors. The ultimate goal is to produce entangled states of two cavity fields, named here as entangled-field Schr\"{o}dinger cats (EFSC).
A Schr\"{o}dinger cat state is a superposition of two distinct coherent states, and our desired state (EFSC) is an entangled state of two such superposed pairs of the two cavity fields \cite{chai1992two,mogilevtsev1996generation,
vitali2000generating}. Such useful states for quantum information processing are experimentally generated using various schemes \cite{van2001entangled,kanari2022two,huang2020generation}. Here, we consider high-quality cavities, guided by dispersive atom-field interaction, where conditional measurements of three atoms will leave behind the desired EFSC for the cavity fields.
The correlation between the two cavity fields is quantified by means of entanglement entropy, that is, the von Neumann entropy of the reduced state. The variation of the entanglement with the cavity field parameters reveals a sophisticated engineering of the desired entanglement of the resulting state. We also riddle out maximally entangled Schr\"{o}dinger cat from a variety of obtained EFSC. A systematic Wigner phase space analysis of the individual reduced state of each cavity manifests quantum interference patterns typical of Schrodinger cat and compass states, even though, due to the entanglement, the state is highly mixed state. We demarcate the parameter regimes for various levels of entanglement, ranging from un-entangled to maximally-entangled ones and display their phase space structure to identify appropriate quantum states for quantum sensing, mediated by entanglement.

The paper is organized as follows. The next section deals with the analytical derivation from the proposed scheme to obtain a family of EFSC. The entanglement of the obtained states is characterized in Sec. III and their variety is delineated through engineering of experimental parameters. Section IV describes the behavior of the states in phase space, along with their nonlocal and spatial interference structures. A brief summary and a future outlook of the work are furnished in the concluding section.

\section{Quantum Circuit of Cavity QED for Generating the EFSC}

\begin{figure*}[htpb]
    \centering
    \includegraphics[width=.67 \textwidth]{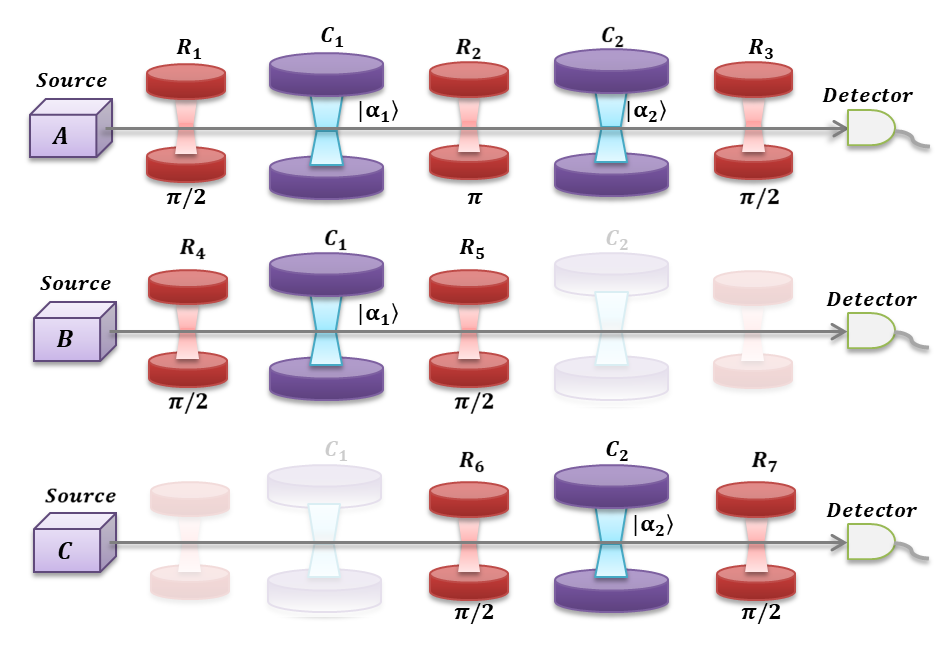}
    \caption{Schematic diagram of a cavity-QED circuit for preparing the EFSC. Three Rydberg atoms, named as $A$, $B$ and $C$, are sent one after another through the two cavity setup. The cavities, labelled as $C_{1}$ and $C_{2}$, are operated under both ``ON" and ``OFF" (faded) conditions and initially prepared with coherent states, $|\alpha_{1}\big\rangle$ and $|\alpha_{2}\big\rangle$, respectively. Ramsey zones are placed on either sides of each cavity with $\pi$ or $\pi/2$ pulses.}
    \label{ch5Schematic}
 \end{figure*}

We consider a two-cavity setup for manipulating three two-level atoms, where each cavity is initially prepared in a coherent state, which will dispersively interact with the atoms. The underlying Hamiltonian is written as per the well-known effective Jaynes Cumming model in the dispersive limit:
\begin{eqnarray}
  H=\hbar\chi(\sigma_{+}\sigma_{-}+a^{\dag}a\sigma_{3}),
\label{JaynesCumming}
\end{eqnarray}
where $\chi$ is the dispersive coupling rate, $\sigma_{+}$ and $\sigma_{-}$ are the raising and lowering operators of the two-level atoms, respectively. Three atoms have been used, consecutively passing through the two cavity setup, where the atomic Rydberg levels, $|g\big\rangle$, $|e\big\rangle$ and $|f\big\rangle$ are the ones mainly interacting with the cavity modes. The level $|f\big\rangle$ is considered away from the other two levels to have dispersive interaction in place. The corresponding atomic inversion operator is given by $\sigma_{3}=|e\big\rangle\big\langle e|-|g\big\rangle\big\langle g|$.
\\
In the dispersive case, the resonant frequency between the two levels is far-detuned from the cavity fundamental frequency. The cavity fields are prepared in coherent states, which are initially independent to each other. There are Ramsey interferometric zones in either sides of the cavities to enable preparing atoms in desired superposition states, mostly choosing a $\pi$ or a $\pi/2$ pulse in our case. The proposed experimental setup for the whole design to manufacture a variety of entangled quantum superposition states is given in Fig. \ref{ch5Schematic}. $C_{1}$ and $C_{2}$ are two microwave cavities, initially contain coherent fields, $|\alpha_{1}\big\rangle$ and $|\alpha_{2}\big\rangle$, respectively. $R_{1}$ to $R_{7}$ are Ramsey interferometric zones with $\pi/2$ pulses, except $R_{2}$ which is guided by a $\pi$ pulse. The three boxes in Fig. \ref{ch5Schematic} at the beginning of the channels are the sources of atoms, $A$, $B$ and $C$, in a selectively prepared initial state. Using proper Stark shift pulses, atom $B$ can be made noninteracting with cavity $C_{2}$, while atom $C$ can be made noninteracting with cavity $C_{1}$ (shown by the faded cavities in Fig. \ref{ch5Schematic}). There are detectors for measuring the atoms in desired combination of tripartite states. In the atom-cavity dynamics described in the following section, we will neglect any decay process, both for atoms and for cavities, by exploiting the exceptional long lifetimes of circular Rydberg states and the fact that the cavities possess a very high quality factor.

\begin{table*}[htpb]
\begin{tabular}{|c||c|}
\hline
    \multicolumn{1}{|c||}{Atomic state } & \multicolumn{1}{c|}{EFSC}  \\ \hline
             \;\;$|g_{1}\big\rangle |g_{2}\big\rangle |g_{3}\big\rangle$\;\; & \;\;$|\psi_{1}\big\rangle  = \frac{N_{1}}{8}\bigg[\Bigr(|\alpha_{1}e^{2i\chi_{1}t}\big\rangle + e^{-i\chi_{1}t}|\alpha_{1}\big\rangle\Bigr)\Bigr(|\alpha_{2}e^{i\chi_{2}t}
             \big\rangle + e^{-i\chi_{2}t}|\alpha_{2}e^{-i\chi_{2}t}\big\rangle\Bigr)$\;\;  \\
             \;\; & \;\; $- \Bigr(|\alpha_{1}e^{i\chi_{1}t}\big\rangle + e^{-i\chi_{1}t}|\alpha_{1}e^{-i\chi_{1}t}\big\rangle\Bigr)
             \Bigr(|\alpha_{2}e^{2i\chi_{2}t}\big\rangle +e^{-i\chi_{2}t}|\alpha_{2}\big\rangle\Bigr)\bigg] $ \;\;\\ \hline
             \;\;$|g_{1}\big\rangle |e_{2}\big\rangle |g_{3}\big\rangle$\;\; & \;\;$|\psi_{2}\big\rangle  = \frac{N_{2}e^{i\delta}}{8}\bigg[\Bigr(|\alpha_{1}e^{2i\chi_{1}t}\big\rangle - e^{-i\chi_{1}t}|\alpha_{1}\big\rangle\Bigr)\Bigr(|\alpha_{2}e^{i\chi_{2}t}
             \big\rangle + e^{-i\chi_{2}t}|\alpha_{2}e^{-i\chi_{2}t}\big\rangle\Bigr)$\;\;  \\
             \;\; & \;\; $- \Bigr(|\alpha_{1}e^{i\chi_{1}t}\big\rangle - e^{-i\chi_{1}t}|\alpha_{1}e^{-i\chi_{1}t}\big\rangle\Bigr)
             \Bigr(|\alpha_{2}e^{2i\chi_{2}t}\big\rangle +e^{-i\chi_{2}t}|\alpha_{2}\big\rangle\Bigr)\bigg] $ \;\;\\ \hline
             \;\;$|f_{1}\big\rangle |g_{2}\big\rangle |g_{3}\big\rangle$\;\; & \;\;$|\psi_{3}\big\rangle  = \frac{-N_{3}e^{i\delta}}{8}\bigg[\Bigr(|\alpha_{1}e^{2i\chi_{1}t}\big\rangle + e^{-i\chi_{1}t}|\alpha_{1}\big\rangle\Bigr)\Bigr(|\alpha_{2}e^{i\chi_{2}t}
             \big\rangle + e^{-i\chi_{2}t}|\alpha_{2}e^{-i\chi_{2}t}\big\rangle\Bigr)$\;\;  \\
             \;\; & \;\; $+ \Bigr(|\alpha_{1}e^{i\chi_{1}t}\big\rangle + e^{-i\chi_{1}t}|\alpha_{1}e^{-i\chi_{1}t}\big\rangle\Bigr)
             \Bigr(|\alpha_{2}e^{2i\chi_{2}t}\big\rangle +e^{-i\chi_{2}t}|\alpha_{2}\big\rangle\Bigr)\bigg] $ \;\;\\ \hline
             \;\;$|f_{1}\big\rangle |e_{2}\big\rangle |g_{3}\big\rangle$\;\; & \;\;$|\psi_{4}\big\rangle  = \frac{-N_{4}e^{i2\delta}}{8}\bigg[\Bigr(|\alpha_{1}e^{2i\chi_{1}t}\big\rangle -e^{-i\chi_{1}t}|\alpha_{1}\big\rangle\Bigr)\Bigr(|\alpha_{2}e^{i\chi_{2}t}
             \big\rangle + e^{-i\chi_{2}t}|\alpha_{2}e^{-i\chi_{2}t}\big\rangle\Bigr)$\;\;  \\
             \;\; & \;\; $+ \Bigr(|\alpha_{1}e^{i\chi_{1}t}\big\rangle - e^{-i\chi_{1}t}|\alpha_{1}e^{-i\chi_{1}t}\big\rangle\Bigr)
             \Bigr(|\alpha_{2}e^{2i\chi_{2}t}\big\rangle +e^{-i\chi_{2}t}|\alpha_{2}\big\rangle\Bigr)\bigg] $ \;\;\\ \hline
             \;\;$|g_{1}\big\rangle |g_{2}\big\rangle |e_{3}\big\rangle$\;\; & \;\;$|\psi_{5}\big\rangle  = \frac{N_{5}e^{i\delta}}{8}\bigg[\Bigr(|\alpha_{1}e^{2i\chi_{1}t}\big\rangle + e^{-i\chi_{1}t}|\alpha_{1}\big\rangle\Bigr)\Bigr(|\alpha_{2}e^{i\chi_{2}t}
             \big\rangle - e^{-i\chi_{2}t}|\alpha_{2}e^{-i\chi_{2}t}\big\rangle\Bigr)$\;\;  \\
             \;\; & \;\; $- \Bigr(|\alpha_{1}e^{i\chi_{1}t}\big\rangle + e^{-i\chi_{1}t}|\alpha_{1}e^{-i\chi_{1}t}\big\rangle\Bigr)
             \Bigr(|\alpha_{2}e^{2i\chi_{2}t}\big\rangle - e^{-i\chi_{2}t}|\alpha_{2}\big\rangle\Bigr)\bigg] $ \;\;\\ \hline
             \;\;$|g_{1}\big\rangle |e_{2}\big\rangle |e_{3}\big\rangle$\;\; & \;\;$|\psi_{6}\big\rangle  = \frac{N_{6}e^{i2\delta}}{8}\bigg[\Bigr(|\alpha_{1}e^{2i\chi_{1}t}\big\rangle -e^{-i\chi_{1}t}|\alpha_{1}\big\rangle\Bigr)\Bigr(|\alpha_{2}e^{i\chi_{2}t}
             \big\rangle - e^{-i\chi_{2}t}|\alpha_{2}e^{-i\chi_{2}t}\big\rangle\Bigr)$\;\;  \\
             \;\; & \;\; $- \Bigr(|\alpha_{1}e^{i\chi_{1}t}\big\rangle - e^{-i\chi_{1}t}|\alpha_{1}e^{-i\chi_{1}t}\big\rangle\Bigr)
             \Bigr(|\alpha_{2}e^{2i\chi_{2}t}\big\rangle - e^{-i\chi_{2}t}|\alpha_{2}\big\rangle\Bigr)\bigg] $ \;\;\\ \hline
             \;\;$|f_{1}\big\rangle |g_{2}\big\rangle |e_{3}\big\rangle$\;\; & \;\;$|\psi_{7}\big\rangle  = \frac{-N_{7}e^{i2\delta}}{8}\bigg[\Bigr(|\alpha_{1}e^{2i\chi_{1}t}\big\rangle + e^{-i\chi_{1}t}|\alpha_{1}\big\rangle\Bigr)\Bigr(|\alpha_{2}e^{i\chi_{2}t}
             \big\rangle - e^{-i\chi_{2}t}|\alpha_{2}e^{-i\chi_{2}t}\big\rangle\Bigr)$\;\;  \\
             \;\; & \;\; $+ \Bigr(|\alpha_{1}e^{i\chi_{1}t}\big\rangle + e^{-i\chi_{1}t}|\alpha_{1}e^{-i\chi_{1}t}\big\rangle\Bigr)
             \Bigr(|\alpha_{2}e^{2i\chi_{2}t}\big\rangle - e^{-i\chi_{2}t}|\alpha_{2}\big\rangle\Bigr)\bigg] $ \;\;\\ \hline
             \;\;$|f_{1}\big\rangle |e_{2}\big\rangle |e_{3}\big\rangle$\;\; & \;\;$|\psi_{8}\big\rangle  = \frac{-N_{8}e^{i3\delta}}{8}\bigg[\Bigr(|\alpha_{1}e^{2i\chi_{1}t}\big\rangle - e^{-i\chi_{1}t}|\alpha_{1}\big\rangle\Bigr)\Bigr(|\alpha_{2}e^{i\chi_{2}t}
             \big\rangle - e^{-i\chi_{2}t}|\alpha_{2}e^{-i\chi_{2}t}\big\rangle\Bigr)$\;\;  \\
             \;\; & \;\; $+ \Bigr(|\alpha_{1}e^{i\chi_{1}t}\big\rangle - e^{-i\chi_{1}t}|\alpha_{1}e^{-i\chi_{1}t}\big\rangle\Bigr)
             \Bigr(|\alpha_{2}e^{2i\chi_{2}t}\big\rangle - e^{-i\chi_{2}t}|\alpha_{2}\big\rangle\Bigr)\bigg] $ \;\;\\ \hline
\end{tabular}
\caption{Right Column: The obtained entangled mesoscopic superposition state (EFSC), corresponding to each set of projective measurements on the three atoms (left Column), total possibilities being eight. $N_{i}$ $(i=1\; \textmd{to} \;8)$ are the normalization constants. $\delta$ contributes only to the overall phase.}
\label{atomfieldstates}
\end{table*}

\subsection{Step-wise Circuit Operation and State Preparation}

Here, we provide the detailed analytical steps to evaluate the quantum state as per the sequence of operations, given in the cavity QED circuit (Fig.\ref{ch5Schematic}). The suffixes ($1$, $2$, and $3$), used with the ground state $|g\big\rangle$, exited state $|e\big\rangle$ and the auxiliary state $|f\big\rangle$, correspond to atom-A, B and C, respectively. We will also use the notation, $|\Psi^{ABC}_{R_6}\big\rangle$, for representing the resultant quantum state, when atom-A and atom-B have already crossed and atom-C has just passed Ramsey zone $R_6$.
\\
\textbf{Atom-A:} The source (A) emits atom-A upon velocity selection in its ground state $|g_{1}\big\rangle$, which when passes through the first Ramsey zone becomes a superposition state of $|g_{1}\big\rangle$ and an auxiliary state $|f_{1}\big\rangle$, \emph{i.e.},
 \begin{eqnarray}
   |\Psi^A_{R_1}\big\rangle =\frac{1}{\sqrt{2}}(|g_{1}\big\rangle + e^{i\delta}|f_{1}\big\rangle),
 \end{eqnarray}
where $\delta$ is the relative phase, provided by the Ramsey zone, $(R_{1})$. This linearly superposed atom enters into the cavity $(C_1)$ and interacts with its coherent field for which the initial state of atom-cavity system becomes $|\Psi^A_{C_1}(0)\big\rangle = |\Psi^A_{R_1}\big\rangle|\alpha_{1}\big\rangle$. The Jaynes- Cummings Hamiltonian in Eq.(\ref{JaynesCumming}) will apply on atom-A and field ${C_1}$ to produce the quantum state at time $t$ just after the cavity $(C_1)$:
\begin{eqnarray}
|\Psi^A_{C_1}(t)\big\rangle && = e^{\frac{-iHt}{\hslash}}|\Psi^A_{R_1}\big\rangle|\alpha_{1}\big\rangle\nonumber\\
&& = \frac{1}{\sqrt{2}}\Big(|g_{1}\big\rangle|\alpha_{1}e^{i\chi_{1}t}\big\rangle + e^{i\delta}|f_{1}\big\rangle|\alpha_{1}\big\rangle\Big),\nonumber
\label{atomcavt}
\end{eqnarray}
where $\chi_{1}$ is the dispersive coupling rate for cavity-$C_1$. We have used the relations,
$e^{\frac{-iHt}{\hslash}}|g_{1}\big\rangle|\alpha_{1}\big\rangle=|g_{1}\big\rangle|\alpha_{1}e^{i\chi_{1}t}\big\rangle$ and $e^{\frac{-iHt}{\hslash}}|f_{1}\big\rangle|\alpha_{1}\big\rangle = |f_{1}\big\rangle|\alpha_{1}\big\rangle$. It is worth pointing out that the above state in Eq. (\ref{atomcavt}) is an entangled state of atom-A and field ${C_1}$. Neither atom-A, nor $C_1$ have distinguishable quantum state anymore, instead, they share a resultant quantum state of the atom-field system, irrespective of their spatial separation. Then, atom-A continues to move and interacts with the second Ramsey zone $(R_{2})$, guided by a $\pi$-pulse, which transforms the component atomic states as $|g_{1}\big\rangle$ $\rightarrow$ $e^{i\delta}|f_{1}\big\rangle$ and $|f_{1}\big\rangle$ $\rightarrow$ $- e^{-i\delta}|g_{1}\big\rangle$. Such state gets further modified due to the interaction of atom-A and field ${C_2}$, which produces the state
\begin{equation}
|\Psi^A_{C_2}(t)= \frac{1}{\sqrt{2}} \Big(e^{i\delta}|f_{1}\big\rangle|\alpha_{1}e^{i\chi_{1}t}\big\rangle|\alpha_{2}\big\rangle - |g_{1}\big\rangle|\alpha_{1}\big\rangle|\alpha_{2}e^{i\chi_{2}t}\big\rangle\Big).
\label{atomcavtt}
\end{equation}
The final step for Atom-A is a $\pi/2$-pulse in the Ramsey zone $(R_{3})$, which transforms the state as $|g_{1}\big\rangle \rightarrow \frac{1}{\sqrt{2}}(|g_{1}\big\rangle + e^{i\delta}|f_{1}\big\rangle)$ and $|f_{1}\big\rangle \rightarrow \frac{1}{\sqrt{2}}(|g_{1}\big\rangle e^{-i\delta} - |f_{1}\big\rangle)$. Hence, the
entangled state of Eq. (\ref{atomcavtt}) transforms to
\begin{eqnarray}
|\Psi^A_{R_3}\big\rangle = && \frac{1}{2}\Bigr[|g_{1}\big\rangle \Big(|\alpha_{1}e^{i\chi_{1}t}\big\rangle|\alpha_{2}\big\rangle - |\alpha_{1}\big\rangle|\alpha_{2}e^{i\chi_{2}t}\big\rangle\Big)- \nonumber\\ && e^{i\delta}|f_{1}\Big(|\alpha_{1}e^{i\chi_{1}t}\big\rangle|\alpha_{2}\big\rangle + |\alpha_{1}\big\rangle|\alpha_{2}e^{i\chi_{2}t}\big\rangle\Big)\Bigr].
\label{ch5waveA}
\end{eqnarray}
\textbf{Atom-B:} In the second stage, atom-B is sent through the two-cavity system, where only cavity-$C_1$ interacts with the atom. Prior to that, the atom is prepared in a superposition state by passing through the Ramsey zone, $R_{4}$, guided by a $\pi/2$-pulse and the atomic state before entering into the cavity becomes $ \frac{1}{\sqrt{2}}(|g_{2}\big\rangle + e^{i\delta}|e_{2}\big\rangle)$ with a relative phase $\delta$. This atom dispersively interacts with field in ${C_1}$ (already sharing the state $|\Psi^A_{R_3}\big\rangle$ of Eq. (\ref{ch5waveA})) through the Jaynes-Cummings Hamiltonian to obtain the state of the system as
\begin{eqnarray*}
|\Psi^{AB}_{C_1}\big\rangle = && \frac{1}{2\sqrt{2}}\Bigr[|g_{1}\big\rangle \tilde{\alpha_1} - e^{i\delta} |f_{1}\big\rangle \tilde{\alpha_2}\Bigr]|g_{2}\big\rangle + \\
&& \frac{1}{2\sqrt{2}}e^{i\delta-i\chi_{1} t}\Bigr[|g_{1}\big\rangle \tilde{\alpha_3} - e^{i\delta} |f_{1}\big\rangle \tilde{\alpha_4}\Bigr]|e_{2}\big\rangle,
\label{}
\end{eqnarray*}
where the cavity field states are collected in the terms $\tilde{\alpha_1}$, $\tilde{\alpha_2}$, $\tilde{\alpha_3}$ and $\tilde{\alpha_4}$, which are the entangled states of the two cavity fields detailed in Appendix-I.

This state gets further mixed up when atom-B passes through $R_{5}$ with a $\pi/2$-pulse and the resultant entangled quantum state of the system formed by atoms A and B and the fields of cavities ${C_1}$ and ${C_2}$, becomes
\begin{widetext}
\begin{eqnarray}
|\Psi^{AB}_{R_{5}}\big\rangle = && \frac{1}{4}\Bigr[|g_{1}\big\rangle |g_{2}\big\rangle \Bigr(\tilde{\alpha_1}+e^{-i\chi_{1}t}\tilde{\alpha_3}\Bigr)+ e^{i\delta}|g_{1}\big\rangle|e_{2}\big\rangle\Bigr(\tilde{\alpha_1}-e^{-i\chi_{1}t}\tilde{\alpha_3}\Bigr)\nonumber\\
&& -e^{i\delta}|f_{1}\big\rangle|g_{2}\big\rangle\Bigr(\tilde{\alpha_2} + e^{-i\chi_{1}t}\tilde{\alpha_4}\Bigr)- e^{2 i\delta}|f_{1}\big\rangle|e_{2}\big\rangle\Bigr(\tilde{\alpha_2} - e^{-i\chi_{1}t}\tilde{\alpha_4}\Bigr)\Bigr].
\label{}
\end{eqnarray}
\end{widetext}

\textbf{Atom-C:} The third atom (C) is made to interact with cavity-$C_2$, Ramsey zones $R_6$ and $R_7$ (both with a $\pi/2$-pulse). Instead of writing the lengthy expression here, we would like to write the final state of the whole setup after Ramsey zone $R_7$ as Eq.(\ref{ch5finalwave}) in Appendix-I. The final state has $64$ terms in total, where the accumulated field coefficients, different from the earlier ones, are ${\alpha_{i}}'$ and ${\alpha_{i}}''$ with $i=1,\;2,\;3,\;4$. These are also entangled states of the phase-modulated two cavity fields and are given in Appendix-I.
\\
In Eq.(\ref{ch5finalwave}), for three two-level atoms, there are eight combinations of atomic states. Each combination has the associated field coefficient as an entangled state of cavities $C_1$ and $C_2$, with eight of their tensor product terms. Conditional measurements on the three atoms generate the desired EFSC for the state of the two cavities, which does not depend on $\delta$, while depending upon $\chi_1 t$, $\chi_2 t$, and the coherent field amplitudes $\alpha_1$ and $\alpha_2$. Hence, it is apparent that Eq.(\ref{ch5finalwave}) offers a family of EFSC, which can be produced by tuning these physical parameters in the experiment. The joint atomic measurements and their corresponding conditionally generated EFSC states are listed in Table-\ref{atomfieldstates}.

There are eight measurement outcomes in total, each of which exhibits entanglement between two quantum superposition states of the cavity fields. We underline that the cavity field components of the nonseparable state are individually in the form of a Schr\"{o}dinger cat state. When one cavity is measured in a Schr\"{o}dinger cat state, the other will also be projected onto another Schr\"{o}dinger cat state. The separability between the constituent coherent states of each cat depends on the tuning parameters, $\chi_{1},\; \chi_{2},\;\alpha_{1}\; \textmd{and}\; \alpha_{2}$.
\\
\\
There are two types of generic mesoscopic superposition: i) superposition between $|\alpha\big\rangle$ and $|e^{2i\chi t}\alpha\big\rangle$; and ii) superposition between $|e^{-i\chi t}\alpha\big\rangle$ and $|e^{i\chi t}\alpha\big\rangle$. In both cases, the coherent states are separated in phase space by $\sqrt{2}$ times the absolute value of the difference of their magnitudes. Hence, the minimum difference happens for $\chi t= n\pi$ with integer $n$, which provides an untangled state because it is equivalent to having no interaction inside the cavities, whereas the maximum separation is obtained for $\chi t=\pi/2$. The latter case produces an entanglement of two cat states placed along orthogonal directions in phase space. It would be interesting to find the quantum states for the maximum separation for the Schr\"{o}dinger cats of both the cavities, \emph{i.e.,} $\chi_1 t=\chi_2 t=\pi/2$. We use this specific case for all the states in Table \ref{atomfieldstates} and obtain eight EFSC with maximized phase-space distance. We choose this option, corresponding to compass-like states useful for optimal displacement quantum sensing, and which are listed in Appendix-I from Eq. (\ref{$g1g2g3$}) to Eq. (\ref{$f1e2e3$}).
It is worth noticing that, out of the eight entangled field states, four states $\psi_{1},\;\psi_{3},\;\psi_{6}\;\textmd{and}\;\psi_{8}$ take the form of Bell-like states of the form $\Bigr(|\phi_1\rangle |\phi_2\rangle \pm |\phi_2\rangle |\phi_1\rangle\Bigr)$. The remaining four states can be also entangled and any conditional generation of these states will leave the state of the cavities in a maximally separated Schr\"{o}dinger cat state, either in position (real cat) or in momentum (imaginary cat). However, they do not take the form of a quasi-Bell state. A quasi-Bell state involving two cat states, one along the real and the other along the imaginary axis of the phase space, resembles compass states, i.e., superpositions of four equally spaced coherent states, whose interference properties are known to manifest structures with dimension smaller than Planck constant $(\hbar)$, as first noticed by Zurek \cite{zurek2001sub}.
We now study the entanglement of the conditionally generated states, and we will then study the phase space properties in Sec. IV.

\begin{figure*}[htb]
    \centering
    \includegraphics[width=.99 \textwidth]{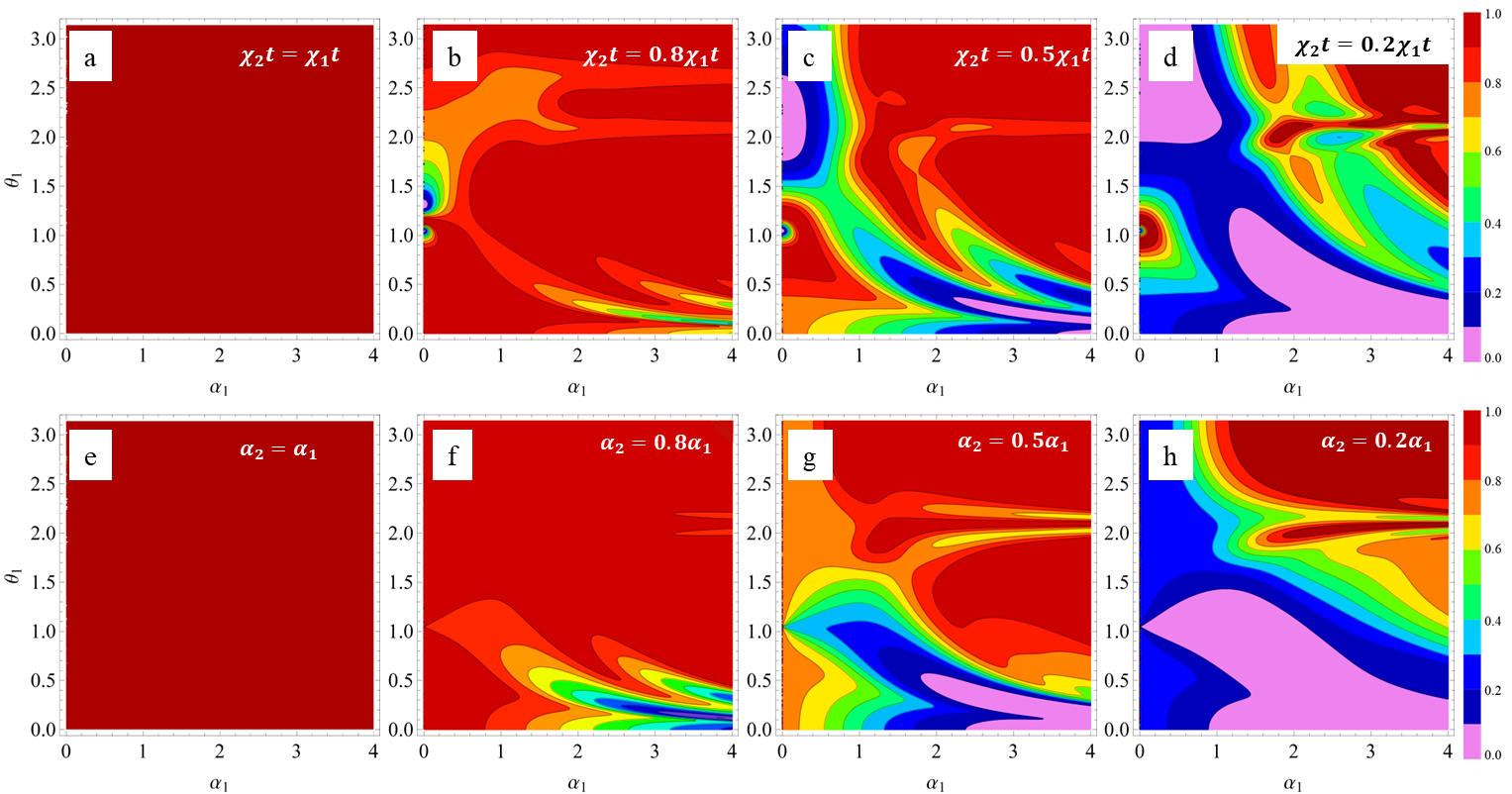}
    \caption{von Neumann entropy of the quantum state of the first subsystem upon measuring the atoms in $|g_{1}\rangle |g_{2}\rangle |g_{3}\rangle$ at various parameter regimes. Upper row corresponds to $\alpha_{1}=\alpha_{2}$ while $\theta_{1}$ and $\theta_{2}$ are varying. Lower row corresponds to $\theta_{1}=\theta_{2}$ while $\alpha_{1}$ and $\alpha_{2}$ are varying. The region shaded in dark-red corresponds to $E \approx 1$, representing the region of maximal entanglement, and the region in pink corresponds to no entanglement $E \approx 0$. We note that when $\alpha_{1}=\alpha_{2}$ and $\theta_{1}=\theta_{2}$ (Figs. (a) and (e)) we have maximum entanglement for all the parameter values.}
    \label{g1g2g3EntropyAll}
 \end{figure*}
\begin{figure*}[htb]
    \centering
    \includegraphics[width=.99 \textwidth]{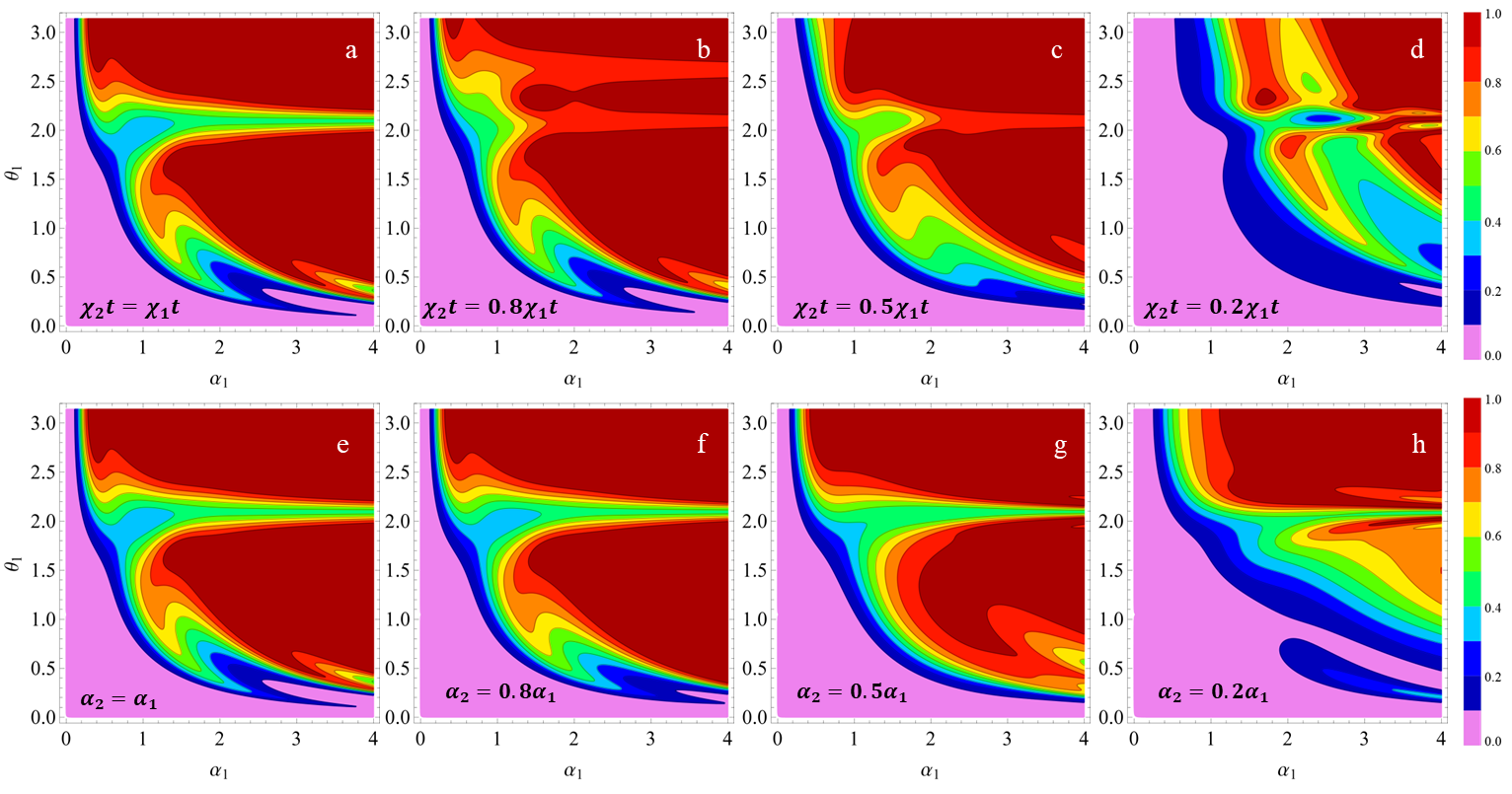}
    \caption{von Neumann entropy of the quantum state of the first subsystem upon measuring the atoms in $|f_{1}\rangle |g_{2}\rangle |e_{3}\rangle$ at various parameter regimes. Upper row corresponds to $\alpha_{1}=\alpha_{2}$ while $\theta_{1}$ and $\theta_{2}$ are varying. Lower row corresponds to $\theta_{1}=\theta_{2}$ while $\alpha_{1}$ and $\alpha_{2}$ are varying. The region shaded in dark-red corresponds to $E \approx 1$, representing the region of maximal entanglement and the region in pink corresponds to no entanglement $E \approx 0$. Fig (a) and (e) depict the maximally entangled region for all parameter values, but satisfying $\alpha_{1}=\alpha_{2}$ and $\theta_{1}=\theta_{2}$.}
    \label{f1g2g3EntropyAll}
\end{figure*}
\begin{figure}[htpb]
    \centering
    \includegraphics[width=.45 \textwidth]{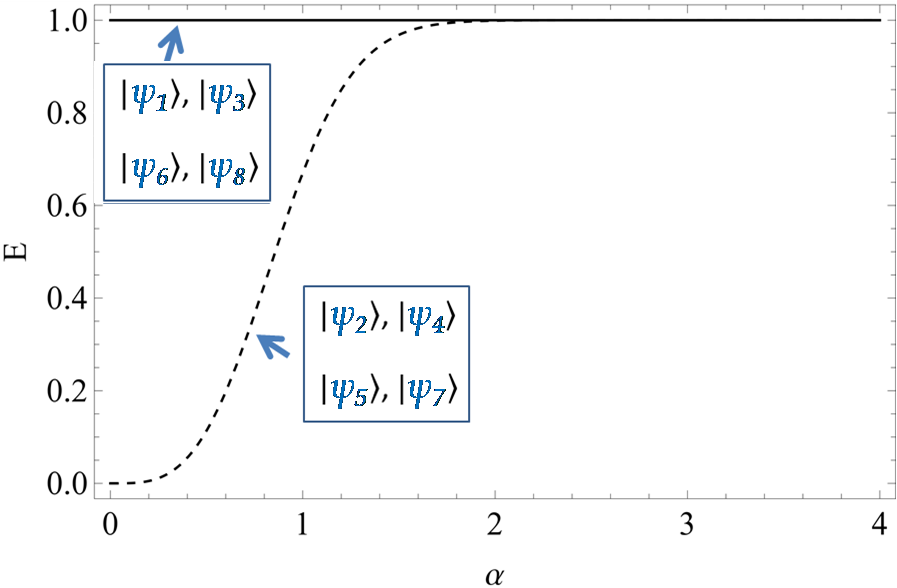}
    \caption{von Neumann entropy of the eight quantum states in the most nonlocal situation, $\chi_{1}t=\chi_{2}t=\pi /2$ and $\alpha_1=\alpha_2=\alpha$. Solid horizontal line corresponds to states, $|\psi_{j}\big\rangle$ : \{$j=1,3,6$ and $8$\}, having one ebit of entanglement and suitable for teleportation. Dashed curve refers to the entanglement for the states $|\psi_{j}\big\rangle$ : \{$j=2,4,5$ and $7$\}, which saturates to unity only for large enough values of $\alpha$. Only for large amplitudes the two single mode states becomes quasi-orthogonal and the von-Neumann entropy achieves its maximum value. }
    \label{ch5entropy}
\end{figure}

\section{Quantitative Estimate of the Entanglement}

As we have seen in the previous section, the eight conditionally generated quantum states are entangled Schr\"{o}dinger cat states of the two cavities. Four of them match the mathematical form of Bell states \cite{hirota2001entangled,hirota2002entangled,ahmad2006entangled,
zeng2007nonclassical,bukhari2011two,hirota2017error,kato2015quasi}.
If the involved states are orthogonal, then the quasi-Bell states can be seen as maximally entangled Bell states.
\\
\\
Now, it is worthwhile to quantify the entanglement of these states, which we do here by using the von-Neumann entropy of the reduced state, which is a proper entanglement measure for bipartite pure states \cite{Horodecki}. The von-Neumann entropy of the $j$-th cavity field is given by $E=-Tr_{3-j}[\rho_{j}\log\rho_{j}]$, which can be rewritten as $E=-\sum_{i}\lambda_{i}\log\lambda_{i}$, where $\lambda_{i}$ are the nonzero eigenvalues of the reduced density matrix $\rho_j$ of the $j$-th subsystem. We have analytically evaluated this for all the obtained quantum states, but we show that here for $|\psi_{1}\big\rangle$ only. The other entropy calculations can be performed in a similar manner. The selected state $|\psi_{1}\big\rangle$ is the state conditionally generated when we detect $|g_{1}\rangle |g_{2}\rangle |g_{3}\rangle$, and is rewritten as
\begin{widetext}
\begin{eqnarray}
|\psi_{1}\big\rangle & = N_{1}\bigg[\Bigr(|\alpha_{1}e^{2i\chi_{1}t}\big\rangle + e^{-i\chi_{1}t}|\alpha_{1}\big\rangle\Bigr)\Bigr(|\alpha_{2}e^{i\chi_{2}t}\big\rangle + e^{-i\chi_{2}t}|\alpha_{2}e^{-i\chi_{2}t}\big\rangle\Bigr)\nonumber\\
& - \Bigr(|\alpha_{1}e^{i\chi_{1}t}\big\rangle + e^{-i\chi_{1}t}|\alpha_{1}e^{-i\chi_{1}t}\big\rangle\Bigr)\Bigr(|\alpha_{2}e^{2i\chi_{2}t}\big\rangle +e^{-i\chi_{2}t}|\alpha_{2}\big\rangle\Bigr)\bigg],
\label{$Geng1g2g3$}
\end{eqnarray}
\end{widetext}
Writing $|\psi_{1}\big\rangle = N_{1}\big[|\mu_{1}\rangle|\nu_{2}\rangle-
|\nu_{1}\rangle|\mu_{2}\rangle\big]$, the corresponding bases modulo normalization become
\begin{eqnarray*}
|\mu_{1}\rangle  = & |\alpha_{1}e^{2i\chi_{1}t}\big\rangle + e^{-i\chi_{1}t}|\alpha_{1}\big\rangle,\nonumber\\
|\nu_{2}\rangle  = & |\alpha_{2}e^{i\chi_{2}t}\big\rangle + e^{-i\chi_{2}t}|\alpha_{2}e^{-i\chi_{2}t}\big\rangle, \nonumber\\
|\nu_{1}\rangle  = & |\alpha_{1}e^{i\chi_{1}t}\big\rangle + e^{-i\chi_{1}t}|\alpha_{1}e^{-i\chi_{1}t}\big\rangle, \nonumber\\
|\mu_{2}\rangle  = & |\alpha_{2}e^{2i\chi_{2}t}\big\rangle + e^{-i\chi_{2}t}|\alpha_{2}\big\rangle.\nonumber
\label{}
\end{eqnarray*}
The transformation from the non-orthogonal basis to orthogonal basis is done for both the cavity modes:
\begin{eqnarray}
{|0\rangle}_{1} = & |\mu_{1}\rangle, \nonumber\\
{|1\rangle}_{1} = & \Big(|\nu_{1}\rangle-p_{1}{|0\rangle}_{1}\Big)/q_{1}, \nonumber
\end{eqnarray}
and
\begin{eqnarray}
{|0\rangle}_{2} = & |\mu_{2}\rangle, \nonumber\\
{|1\rangle}_{2} = & \Big(|\nu_{2}\rangle-p_{2}{|0\rangle}_{2}\Big)/q_{2}, \nonumber
\end{eqnarray}
respectively. Here, $p_{1}=\langle\mu_{1}|\nu_{1}\rangle$, $q_{1}=\sqrt{1-|p_{1}|^{2}}$, $p_{2}=\langle\mu_{2}|\nu_{2}\rangle$, and $q_{2}=\sqrt{1-|p_{2}|^{2}}$. The quantum state can be rewritten as
\begin{equation}
|\psi_{1}\big\rangle = N_{1}\big[(p_{2}-p_{1}){|0\rangle}_{1}{|0\rangle}_{2}+
q_{2}{|0\rangle}_{1}{|1\rangle}_{2}-
q_{1}{|1\rangle}_{1}{|0\rangle}_{2}\big],
\label{orthostate}
\end{equation}
where $N_1$ is the normalization factor guaranteeing that $|N_{1}|^{2}\big[|p_{2}-p_{1}|^{2}+|q_{2}|^{2}+
|q_{1}|^{2}\big]=1$. As we want to find the entropy of mode 1, the corresponding reduced density matrix is evaluated by tracing out mode 2: $\rho_{1}=Tr_{2}|\psi_{1}\big\rangle \big\langle\psi_{1}|$. The reduced density matrix of mode 1 is written as
\[
\rho_{1}=
  \begin{bmatrix}
    |N_{1}|^{2}\big[|p_{2}-p_{1}|^{2}+|q_{2}|^{2}\big] && -|N_{1}|^{2}(p_{2}-p_{1})q^{\ast}_{1}  \\ \\
    -|N_{1}|^{2}(p_{2}-p_{1})^{\ast}q_{1} && |N_{1}|^{2}|q_{1}|^{2}
  \end{bmatrix},
\]
the determinant of which is $\textmd{det} (\rho_{1})= |N_{1}|^{4}|q_{1}|^{2}|q_{2}|^{2}$ and the eigenvalues are
\begin{eqnarray}
\lambda_{\pm}=\frac{1}{2}\Big[1\pm \sqrt{1-4\; \textmd{det}(\rho_{1})} \Big].
\label{matrixeigens}
\end{eqnarray}
A similar approach can be used for the other pure conditional states of the two cavities. We now illustrate the behavior of the von Neumann entropy for the measurement outcomes of $|g_{1}\rangle |g_{2}\rangle |g_{3}\rangle$ (quasi-Bell state $|\psi_1\rangle $) and $|f_{1}\rangle |g_{2}\rangle |e_{3}\rangle$ (non quasi-Bell state  $|\psi_7\rangle $) in Fig.\ref{g1g2g3EntropyAll} and Fig.\ref{f1g2g3EntropyAll}, respectively, versus the system parameters. We have defined $\chi t=\theta$ for both modes with corresponding suffixes. The entropy variations are displayed for two tuning parameters of the first cavity mode, $\theta_1$ and $\alpha_1$. However, this is affected by the second cavity mode parameters, $\theta_2$ and $\alpha_2$ due to entanglement. It is quite fascinating to observe that, when the parameters of the two cavity modes are the same, $\theta_1=\theta_2$ and $\alpha_{1}= \alpha_{2}$, in the case of a quasi-Bell state, $|\psi_{1}\big\rangle$, one has maximum entanglement $E=1$ for all physical parameters of both modes (see Fig.\ref{g1g2g3EntropyAll}(a) and (e)). This can be easily understood from the fact that in this case one has $p_1=p_2$ implying $q_1=q_2$, and from Eq. (\ref{orthostate}) one can easily see that the state of the cavities becomes just a Bell state involving the two orthogonal states $|0\rangle$ and $|1\rangle$.

On the contrary, deviating from the condition of equal parameters results in a nontrivial variation of the entanglement and even to its disappearance ($E \approx 0$) at some parameter values (see Fig.\ref{g1g2g3EntropyAll}(b)-(d) and (f)-(h)). The upper row of Fig.\ref{g1g2g3EntropyAll} refers to the variation of the ratio $\theta_2/\theta_1$ while keeping $\alpha_1=\alpha_2$, while the lower row refers to the variation of the ratio $\alpha_2/\alpha_1$, by keeping $\theta_1=\theta_2$. The contours are used in such a way that the dark-red color corresponds to the region of very large entanglement with entropy $>0.9$ and pink color shows the region of almost no-entanglement with entropy $<0.1$. The region of no-entanglement gets gradually inhabited at the cost of reduction of the area for maximal entanglement when the ratio, $\theta_2/\theta_1$ or $\alpha_2/\alpha_1$, gradually decreases. These plots clearly reveal the possibility of choosing a desired amount of entanglement between the cavity modes, by tuning the above interaction parameters and the amplitude of the initial coherent states in the cavities.
\\
We now study the behavior of the entanglement entropy for a non quasi-Bell state, $|\psi_{7}\big\rangle$, and we show the results in Fig.\ref{f1g2g3EntropyAll}. Unlike the previous case, identical interaction and amplitude parameters for the modes, $\theta_1=\theta_2$ and $\alpha_{1}= \alpha_{2}$, does not lead to a maximally entangled scenario for all parameters (see Fig.\ref{f1g2g3EntropyAll}(a) and (e)), but only for a designated parameter domain, showed by dark-red color. At the same time, similarly to the previous case of quasi-Bell state, the parameter region with zero entanglement broadens when the ratio, $\theta_2/\theta_1$ or $\alpha_2/\alpha_1$, decreases from $1.0$ to $0.2$.
\\
\\
We now study the particular situation when $\chi_{1}t=\chi_{2}t=\pi /2$, for which a maximum separation of the two coherent states of the superposition is obtained, and the two cat states are directed along two orthogonal directions in phase space. The entanglement measure through the von Neumann entropy of the obtained eight quantum states in this case are shown in Fig.\ref{ch5entropy}, where two groups are clearly visible. One group is formed by $|\psi_{j}\big\rangle$ : \{$j=1,3,6$ and $8$\} with nonzero eigenvalues of the reduced density matrix ${\lambda^{j}_{1}}= {\lambda^{j}_{2}} = \frac{1}{2}$. The corresponding entanglement is $E(|\psi_{j}\big\rangle)= 1$, which implies that these states are always the maximally entangled state with $1$-ebit of entanglement. These are represented by the solid horizontal line in Fig.\ref{ch5entropy}. On the other hand, two non-zero eigenvalues of the reduced density matrices of the remaining four quantum states, $|\psi_{j}\big\rangle$: \{$j=2,4,5$ and $7$\} are given by
\begin{equation}
{\lambda^{j}_{1}}= \frac{(1-e^{-|\alpha_1|^{2}})^{2}}{2(1+e^{-2 |\alpha_1|^{2}})},\;\;\;\; {\lambda^{j}_{2}}= \frac{(1+e^{-|\alpha_2|^{2}})^{2}}{2(1+e^{-2 |\alpha_2|^{2}})}.
\end{equation}
When these values of ${\lambda^{j}_{1,2}}$ are used within the von-Neumann entropy expression, we obtain $E(|\psi_{j}\big\rangle)\leq 1$, which is represented by the dashed curve in Fig.\ref{ch5entropy}. However, for a larger amplitude $\alpha$ ($> 2$), all the eight obtained states approach one ebit of entanglement, $E=1$. The above observations can also be seen from the proper cuts of the general entropy contour plots of Fig. \ref{g1g2g3EntropyAll}(a) or (e) and Fig. \ref{f1g2g3EntropyAll}(a) or (e), for $\theta_1=\theta_2=\pi /2$.

\begin{figure*}[htb]
    \centering
    \includegraphics[width=.7 \textwidth]{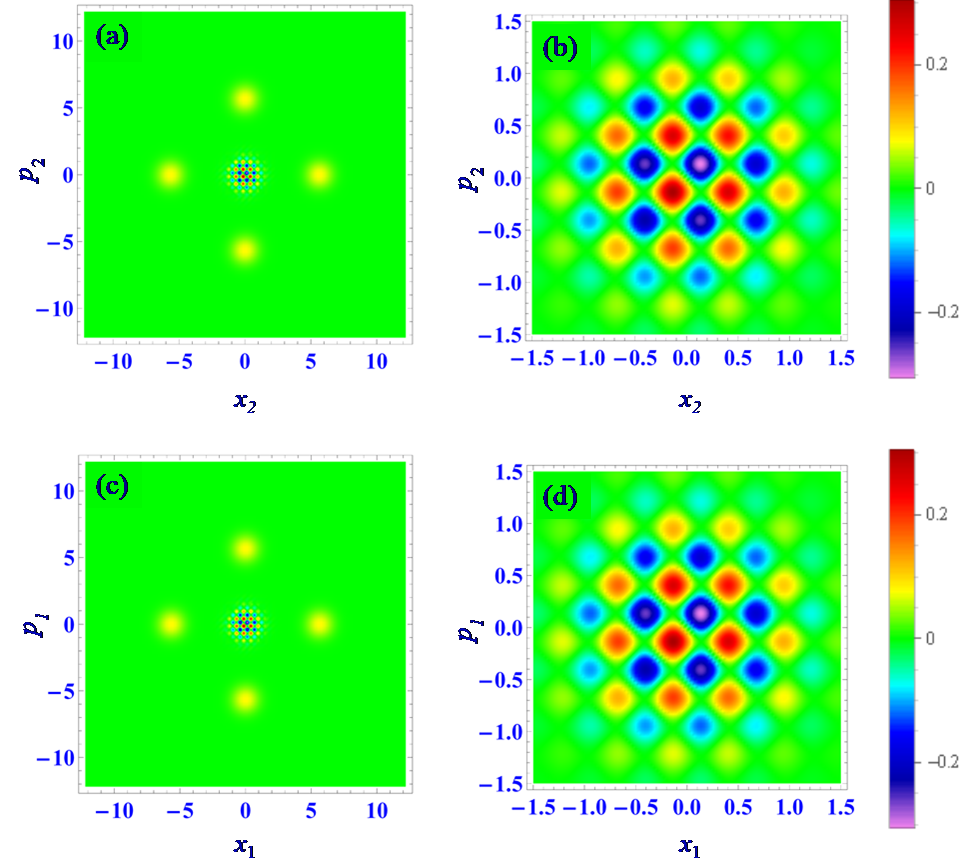}
    \caption{Contour plot of the Wigner function for the state $|\psi_{1}\big\rangle$ in both the phase spaces. For ($x_{1},\;p_{1}$), we integrate the calculated Wigner function over ($x_{2},\;p_{2}$) and vice-versa. Here, we have taken equal phase-space support for both the fields: $\alpha_{1}=\alpha_{2}=4$. Left column ((a) and (c)) shows the reduced Wigner functions of the individual cavity field and the right column ((b) and (d)) display their central interference region.}
    \label{ch5Wg1g2g3}
 \end{figure*}

\begin{figure*}[htb]
    \centering
    \includegraphics[width=.8 \textwidth]{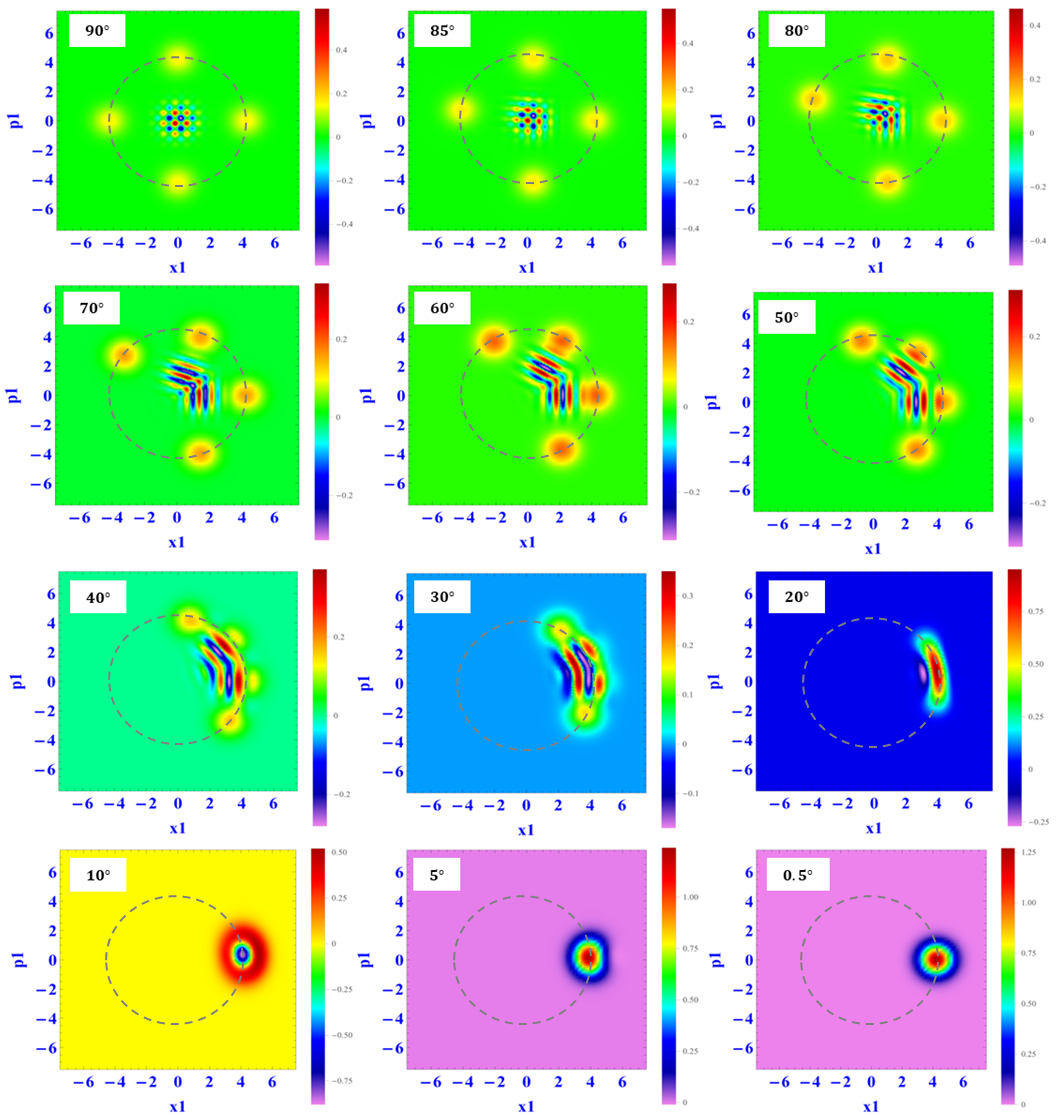}
    \caption{Wigner function in the first phase space ($x_{1},\; p_{1}$) of the entangled field state, obtained from the coefficient $|g_{1}\big\rangle|g_{2}\big\rangle|g_{3}\big\rangle$ in Eq. (\ref{ch5finalwave}). Here, $\chi_{1}t$ and $\chi_{2}t$ both are equal and vary from $90^{\circ}$ to $0.5^{\circ}$. The sub-Planck structures undergo deformation as the constituent coherent states, that make up the system, are rotated and tend to blend together. We consider $\alpha_{1}=\alpha_{2}=3$.}
    \label{ch5g1g2g3_variouswigner}
 \end{figure*}
\begin{figure*}[htb]
    \centering
    \includegraphics[width=.8 \textwidth]{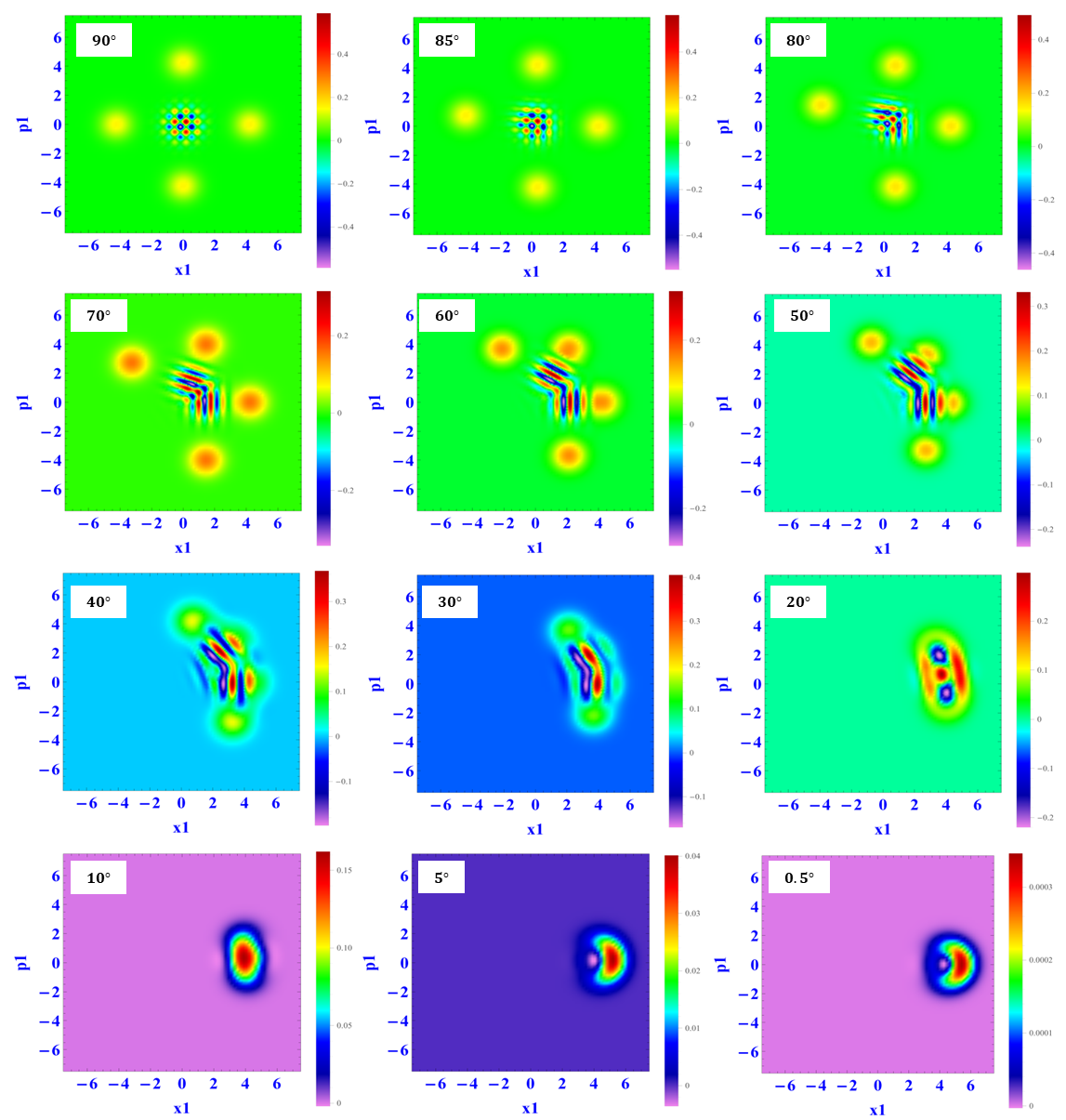}
    \caption{Wigner function in the first phase space ($x_{1},\; p_{1}$) of the entangled field state, obtained from the coefficient $|g_{1}\big\rangle|e_{2}\big\rangle|g_{3}\big\rangle$ in Eq. (\ref{ch5finalwave}). Here, $\chi_{1}t$ and $\chi_{2}t$ both are equal and vary from $90^{\circ}$ to $0.5^{\circ}$. The sub-Planck structures undergo deformation as the constituent coherent states, that make up the system, are rotated and tend to blend together. We consider $\alpha_{1}=\alpha_{2}=3$.}
    \label{ch5g1e2g3_variouswigner}
 \end{figure*}

\section{Phase Space Representation of the EFSC}

In the above sections, we have discussed about the quantum state preparation and their entanglement properties. The generated states possess both the distinct nonclassical features: \emph{a}) entanglement and \emph{b}) mesoscopic quantum superposition. Hence, it is interesting to study their coexistence through phase space Wigner distribution, which has the merit to manifest quantum interference structures of the correlated state.

However, visualizing the phase space of both cavities will requires a four dimensional phase space, which is not possible. Thus, we will visualize one of the field modes, at a time, in its corresponding sub-phase-space by taking the reduced Wigner function upon integrating over the position and momentum variables of the other mode. In this way, we are able to see one cavity field while the two cavities still remain correlated.

Below, we will calculate the Wigner function of a general bipartite state ($|\Psi_{gen}\big\rangle$), which can be projected to any one of the eight listed states of Table \ref{atomfieldstates} modulo normalization:
\begin{widetext}
\begin{eqnarray}
|\Psi_{gen}\big\rangle & =\frac{1}{8}\bigg[A\Bigr(|\alpha_{1}e^{i\theta_{1}}\big\rangle + e^{i\xi_{1}}|-\alpha_{1}e^{i\theta_{1}}\big\rangle\Bigr)\Bigr(|\alpha_{2}e^{i\phi_{2}}\big\rangle + e^{i\zeta_{2}}|-\alpha_{2}e^{i\phi_{2}}\big\rangle\Bigr)\nonumber\\
& + B\Bigr(|\alpha_{1}e^{i\phi_{1}}\big\rangle + e^{i\zeta_{1}}|-\alpha_{1}e^{i\phi_{1}}\big\rangle\Bigr)\Bigr(|\alpha_{1}e^{i\theta_{2}}\big\rangle + e^{i\xi_{2}}|-\alpha_{2}e^{i\theta_{2}}\big\rangle\Bigr)\bigg],
\label{ch5gencatstate}
\end{eqnarray}
\end{widetext}
where $\theta_{1}$, $\theta_{2}$, $\phi_{1}$, $\phi_{2}$, $\xi_{1}$, $\xi_{2}$, $\zeta_{1}$ and $\zeta_{2}$ are the phases with the coherent states and the relative phases between them. $A$ and $B$ are complex coefficients of the constituent entangled pairs. The corresponding two mode Wigner distribution function in combined space with variables, $x_{1}$, $p_{1}$; $x_{2}$, $p_{2}$, can be written as
\begin{eqnarray}
W(x_{1},p_{1};x_{2},p_{2}) && = \frac{1}{\pi^{2}}\int_{-\infty}^{\infty}\Psi_{gen}\big(x_{1}+q,x_{2}+s\big)\nonumber\\
&&\Psi^{\dag}_{gen}\big(x_{1}-q,x_{2}-s\big)e^{-2i(p_{1}q+p_{2}s)}dqds.\nonumber
\label{ch5wigner}
\end{eqnarray}
The quantum state of Eq. (\ref{ch5gencatstate}) is substituted in this expression and a lengthy but straightforward calculation results
\begin{eqnarray}
&& W(x_{1},p_{1};x_{2},p_{2}) = \frac{1}{64{\pi}^2}\exp\{-2\left(x_{1}^{2}+x_{2}^{2}+p_{1}^{2}+p_{2}^{2}\right)\} \nonumber\\
&&\times\bigg[W_{D1}+W_{D2}+W_{OD1}+W_{OD2}\bigg].
\label{ch5totalwigner}
\end{eqnarray}
Here, $W_{D1}$, $W_{D2}$ are diagonal and $W_{OD1}$, $W_{OD2}$ are off-diagonal components of the obtained Wigner function. These components take big expressions and are provided in Appendix-II. This analytical expression of the Wigner function will enable us to visualize the whole family of states with appropriate experimental parameters, subject to its reduction to one of the spaces upon integrating over variables of other space. Below, we will show some of them to reveal their merit towards parameter engineering for entanglement and quantum sensing applications.

\subsection{Wigner Function for a specific case of Largest EFSC}

While visualizing the above Wigner function, one needs to consider specific parameters for the chosen quantum state. The quantum states obtained by Eqs.(\ref{$g1g2g3$} to \ref{$f1e2e3$}) demand specific choices of $\theta_{1}$, $\theta_{2}$, $\phi_{1}$, $\phi_{2}$, $\xi_{1}$, $\xi_{2}$, $\zeta_{1}$, $\zeta_{2}$, $A$ and $B$. In this subsection, we will demonstrate the state obtained in Eq.(\ref{$g1g2g3$}) for instance, which maps the general entangled state by putting $\theta_{1}=0$, $\theta_{2}=0$, $\phi_{1}=\pi/2$, $\phi_{2}=\pi/2$, $\xi_{1}=\pi/2$, $\xi_{2}=\pi/2$, $\zeta_{1}=3\pi/2$, $\zeta_{2}=3\pi/2$, $A=-i$ and $B=i$. This is a Bell state of two Schr\"{o}dinger cats of two independent cavity modes.

Figure \ref{ch5Wg1g2g3} depicts the Wigner function of both the cavity fields separately for $\chi_{1}t=\chi_{2}t=\pi/2$. For first cavity-field, defined by the phase space variables ($x_{1},\;p_{1}$), the corresponding Wigner function is obtained by numerically integrating the total Wigner function over ($x_{2},\;p_{2}$) and vice-versa. The left column of Fig. \ref{ch5Wg1g2g3} displays the states of cavity-$C_1$ (up) and cavity-$C_2$ (down). In this case of equal parameters, the pure state of the two cavities is maximally entangled and it is a Bell state involving the two states $|0\rangle $ and $|1\rangle$ introduced in Sec. III. However, since in this figure we have chosen $\alpha_1=\alpha_2 = 4$, the overlaps $p_1=p_2 \sim 0$ and the reduced state of each cavity is in practice an equal mixture of two cat states directed along the real and imaginary axis of the phase space. In this way the state resembles a compass-state, which is the superposition of four equally spaced coherent states, and show similar interference patterns at the center of the phase space, despite being highly mixed and with large entropy.

\begin{figure*}[htb]
    \centering
    \includegraphics[width=.8 \textwidth]{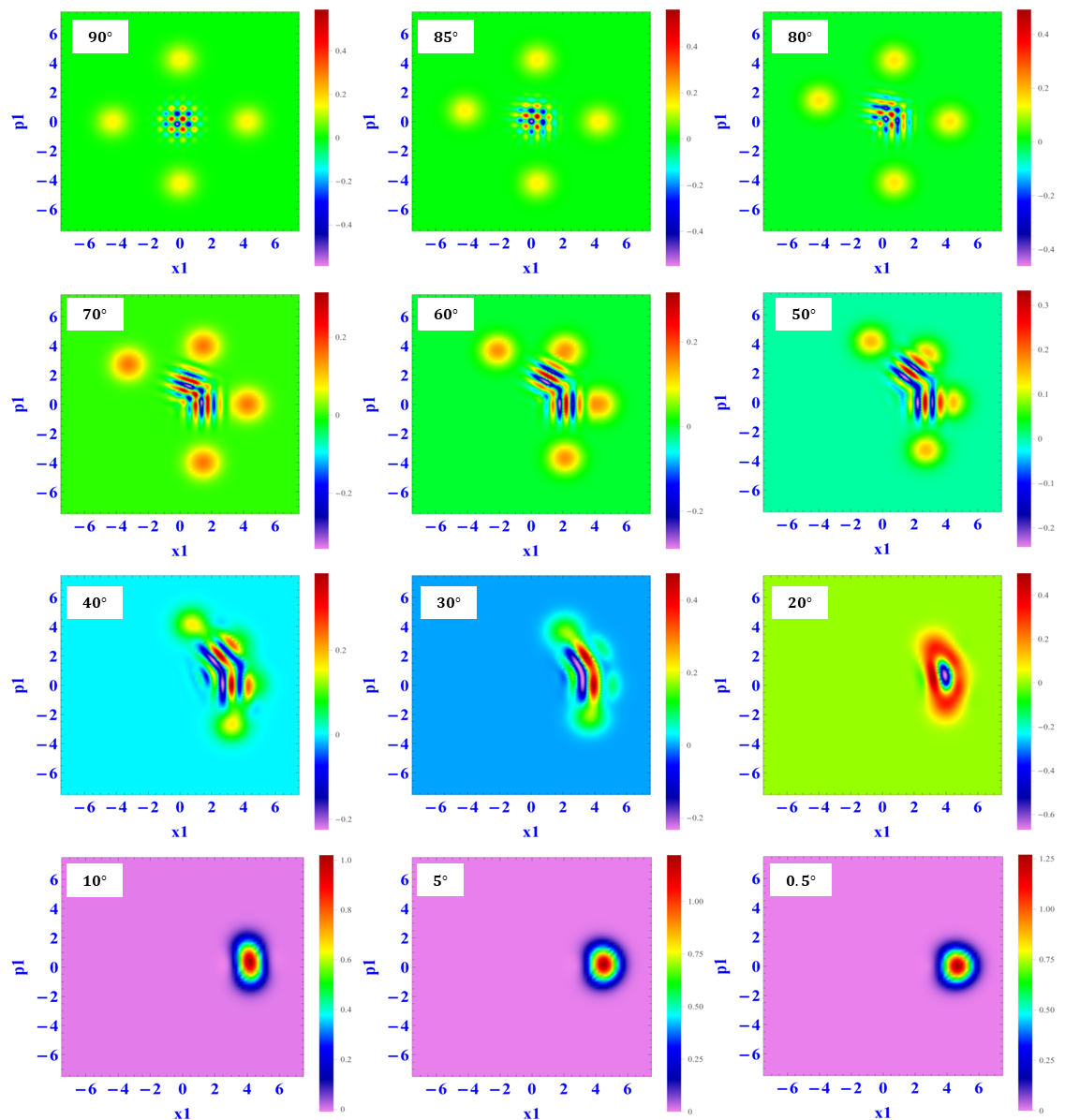}
    \caption{Wigner function in the first phase space ($x_{1},\; p_{1}$) of the entangled field state, obtained from the coefficient $|f_{1}\big\rangle|e_{2}\big\rangle|g_{3}\big\rangle$ in Eq. (\ref{ch5finalwave}). Here, $\chi_{1}t$ and $\chi_{2}t$ both are equal and vary from $90^{\circ}$ to $0.5^{\circ}$. The sub-Planck structures undergo deformation as the constituent coherent states, that make up the system, are rotated and tend to blend together. We consider $\alpha_{1}=\alpha_{2}=3$.}
    \label{ch5f1e2g3_variouswigner}
 \end{figure*}
\begin{figure*}[htb]
    \centering
    \includegraphics[width=.8 \textwidth]{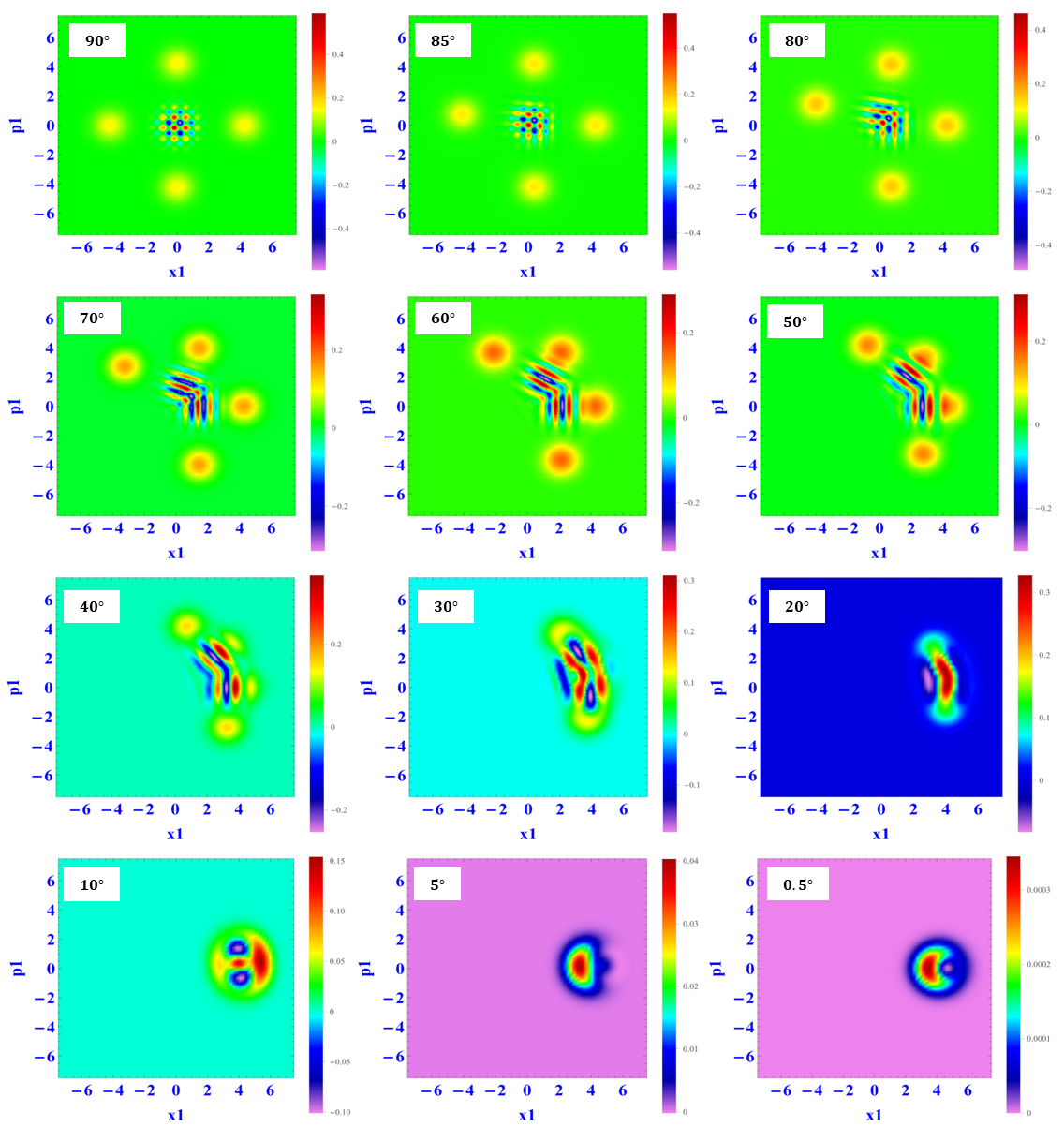}
    \caption{Wigner function in the first phase space ($x_{1},\; p_{1}$) of the entangled field state, obtained from the coefficient $|g_{1}\big\rangle|g_{2}\big\rangle|e_{3}\big\rangle$ in Eq. (\ref{ch5finalwave}). Here, $\chi_{1}t$ and $\chi_{2}t$ both are equal and vary from $90^{\circ}$ to $0.5^{\circ}$. The sub-Planck structures undergo deformation as the constituent coherent states, that make up the system, are rotated and tend to blend together. We consider $\alpha_{1}=\alpha_{2}=3$.}
    \label{ch5g1g2e3_variouswigner}
 \end{figure*}

The right column of Fig. \ref{ch5Wg1g2g3} manifests their corresponding interference structures as a zoomed version of the central region of the phase space. It is quite fascinating to observe such interference structure in the reduced Wigner function of entangled sub-systems, even in a mixed state. These sub-Planck structures, resembling a nice checkerboard pattern, have been extensively studied for quantum sensing. Its sensitivity to environmental decoherence was also reported \cite{jacquod2002decay,wisniacki2003short,pathak2005generation,
ghosh2009sub}. The employment in the measurement of Heisenberg limited sensitivity and quantum parameter estimation in general have been investigated \cite{toscano2006sub,dalvit2006quantum} and compass state in an arrangement with cavity QED has also been proposed \cite{agarwal2004mesoscopic}. Such structures have also become evident in diverse physical contexts: fractional revivals in molecular wave packet \cite{ghosh2006mesoscopic}, entangled cat state \cite{bhatt2008entanglement}, optical fibers \cite{stobinska2008wigner}, Kirkwood-Rihaczek distribution \cite{banerji2007exploring} \emph{etc}. It is remarkable that one can find interference fringes and negative values of the Wigner function even in a highly mixed state, making them also useful for continuous variable quantum computation \cite{Mari2009,Walschaers2023}.

\subsection{Wigner Function of Parameter-engineered EFSC}

Here, we investigate the phase space corresponding to the general quantum state in Eq. (\ref{ch5finalwave}), so that we can incorporate desired tuning to implement entangled state engineering. In all the figures, we have depicted the reduced phase space of the first cavity field, where the parameters of the second cavity also control the state due to their nonseparable mixture. The same analysis is straightforward for the second field in the similar manner. The tunability is incorporated here by changing the coherent field amplitudes; $\chi_{1}t,\;\chi_{2}t$; and all of their asymmetric combinations for both real and imaginary domains of the coherent state parameters. We will demonstrate one of the tunability conditions for quantum state engineering, \textit{i.e.}, by changing $\chi_{1}t$ and $\chi_{2}t$, keeping the field amplitudes fixed ($\alpha_{1}=\alpha_{2}=3$) in our demonstrations. However, it is worth mentioning that, changing $\alpha_{1}$ and $\alpha_{2}$ will alter the phase space support.

Here, we will examine how the constituent coherent states' orientations in one phase space (the first phase space) are changed by decreasing the values of $\chi_{1}t$ and $\chi_{2}t$ from $90^{\circ}$ to $0.5^{\circ}$. The squeezing of the entangled state is reported along with the deformation of their corresponding sub-Planck structures. We have arbitrarily chosen four states from Eq. (\ref{ch5finalwave}), which are the measurement outcomes of the atomic states, $|g_{1}\big\rangle|g_{2}\big\rangle|g_{3}\big\rangle$, $|g_{1}\big\rangle|e_{2}\big\rangle|g_{3}\big\rangle$, $|f_{1}\big\rangle|e_{2}\big\rangle|g_{3}\big\rangle$, and $|g_{1}\big\rangle|g_{2}\big\rangle|e_{3}\big\rangle$ and are depicted in Fig.\ref{ch5g1g2g3_variouswigner}, Fig.\ref{ch5g1e2g3_variouswigner}, Fig.\ref{ch5f1e2g3_variouswigner} and Fig \ref{ch5g1g2e3_variouswigner}, respectively.
\\
\\
Figure \ref{ch5g1g2g3_variouswigner} displays the maximally entangled state of real and imaginary cats in first phase space, where $\chi_{1}t$ and $\chi_{2}t$ vary from $90^{\circ}$ to $0.5^{\circ}$. The state gets gradually deformed from a compass-like state to a Gaussian-like state. However, the four constituent coherent states always remain on a circle of radius $\sqrt{2}|\alpha_1|$, shown by the dotted circle and this is valid for all the figures \ref{ch5g1g2g3_variouswigner}-\ref{ch5g1g2e3_variouswigner}. The state evolution in Fig. \ref{ch5g1e2g3_variouswigner} is quite different from the previous plot. The final plot for $\chi_{1}t=\chi_{2}t=0.5^{\circ}$ is non-Gaussian, but no negative Wigner value. On the contrary, Fig.\ref{ch5f1e2g3_variouswigner} shows a changing nature of the state from $\chi_{1}t=\chi_{2}t=20^{\circ}$ (dip at the middle) to $\chi_{1}t=\chi_{2}t=0.5^{\circ}$ (a Gaussian-like state). The last figure, Fig \ref{ch5g1g2e3_variouswigner}, manifests somewhat an opposite nature from Fig.\ref{ch5g1e2g3_variouswigner} for the lower $\chi_{1}t,\;\chi_{2}t$ values.

In all the above state evolutions, the common features with decreasing $\chi_{1}t,\;\chi_{2}t$ are \textrm{i}) the reduction in the negative phase space region, signifying less non-classical field state, \textrm{ii}) the lowering of the classical action, which is given by the effective phase space support, \textrm{iii}) prominent sub-Planck structures for higher $\chi_{1}t,\;\chi_{2}t$, particularly near to $\chi_{1}t=\chi_{2}t=\pi/2$, and \textrm{iv}) non-negative Wigner function for small $\chi_{1}t,\;\chi_{2}t$, signifying a classical-like state. For $\alpha_{1}=\alpha_{2}=3$, reduced Wigner function of all the eight field states with $\chi_{1}t=\chi_{2}t=\pi/2$ possess maximum entropy (see Fig. \ref{ch5entropy}) and take the shape of a compass-like state (see the first phase space plots of Figs. \ref{ch5g1g2g3_variouswigner}-\ref{ch5g1g2e3_variouswigner}), where the central interference pattern associated to the entangled cat states along two orthogonal directions, persist despite the presence of quantum correlations between them and responsible for the maximum von-Neumann entropy.

\section{Conclusion}

We provide a scheme to produce entanglement between two field Schr\"{o}dinger cats in a cavity QED system. We considered a two cavity setup, where the correlation between the high-Q cavities is established by injecting three Rydberg atoms into them and manipulating their state. A family of entangled states is produced by the scheme, by using the dispersive Jaynes-Cummings model. Entanglement is then quantified by means of the entanglement entropy, that is, the von-Neumann entropy of the reduced state. The generated states show simultaneously two relevant quantum features, Bell-like nonlocal correlations, and also phase-space interference and Wigner negativities, due to the properties of the entangled cat states. These quantum features can be tuned and engineered by changing the dispersive coupling and the amplitude of the coherent states in the cavities. Quantum phase space is thoroughly analyzed through Wigner distribution function, including their state evolutions to address deformation, squeezing and mixing of the entangled states. The entangled cat state generated here could be potentially used for quantum technology applications such as quantum sensing of displacements, or continuous variable quantum computation.

\section{Acknowledgement}
UR acknowledges the support from the project (CRG/2022/007467, SERB, India); DV acknowledges the support of PNRR
MUR project PE0000023-NQSTI (Italy).

\vskip 2cm

\section{Appendix-I}

As mentioned, atom-B after passing through $R_{5}$ (a $\pi/2$-pulse) generates a combined entangled state of $atom_A$-$atom_B$-$field_{C_1}$-$field_{C_2}$, where none of the atoms and fields are separable:
\begin{widetext}
\begin{eqnarray}
|\Psi^{AB}_{R_{5}}\big\rangle = && \frac{1}{4}\Bigr[|g_{1}\big\rangle |g_{2}\big\rangle \Bigr(\tilde{\alpha_1}+e^{-i\chi_{1}t}\tilde{\alpha_3}\Bigr)+ e^{i\delta}|g_{1}\big\rangle|e_{2}\big\rangle\Bigr(\tilde{\alpha_1}-e^{-i\chi_{1}t}\tilde{\alpha_3}\Bigr)\nonumber\\
&& -e^{i\delta}|f_{1}\big\rangle|g_{2}\big\rangle\Bigr(\tilde{\alpha_2} + e^{-i\chi_{1}t}\tilde{\alpha_4}\Bigr)- e^{2 i\delta}|f_{1}\big\rangle|e_{2}\big\rangle\Bigr(\tilde{\alpha_2} - e^{-i\chi_{1}t}\tilde{\alpha_4}\Bigr)\Bigr].
\nonumber
\end{eqnarray}
\end{widetext}
Here, the field entangled pairs are collected in the following terms:
\begin{eqnarray*}
\tilde{\alpha_1} && = \Bigr\{|\alpha_{1}e^{2i\chi_{1}t}\big\rangle|\alpha_{2}\big\rangle - |\alpha_{1}e^{i\chi_{1}t}\big\rangle|\alpha_{2}e^{i\chi_{2}t}\big\rangle\Bigr\}, \nonumber\\
\tilde{\alpha_2} && = \Bigr\{|\alpha_{1}e^{2i\chi_{1}t}\big\rangle|\alpha_{2}\big\rangle + |\alpha_{1}e^{i\chi_{1}t}\big\rangle|\alpha_{2}e^{i\chi_{2}t}\big\rangle\Bigr\},\nonumber\\
\tilde{\alpha_3} && = \Bigr\{|\alpha_{1}\big\rangle|\alpha_{2}\big\rangle - |\alpha_{1}e^{-i\chi_{1}t}\big\rangle|\alpha_{2}e^{i\chi_{2}t}\big\rangle\Bigr\},\nonumber\\
\tilde{\alpha_4} && = \Bigr\{|\alpha_{1}\big\rangle|\alpha_{2}\big\rangle - |\alpha_{1}e^{-i\chi_{1}t}\big\rangle|\alpha_{2}e^{i\chi_{2}t}\big\rangle\Bigr\}.
\end{eqnarray*}
Cavity-$C_1$ does not interact with atom-C, but cavity-$C_2$, Ramsey zones $R_6$ and $R_7$ (both as $\pi/2$-pulse) operate. Using the standard operations described earlier, one will be able to derive the final quantum state of the whole setup as follows
\begin{widetext}
\begin{eqnarray}
|\Psi_{Final}\big\rangle && = \frac{1}{8}\Bigg[|g_{1}\big\rangle |g_{2}\big\rangle |g_{3}\big\rangle \Bigr\{ {\alpha_{1}}' + e^{-i\chi_{1}t}{\alpha_{3}}' + {\alpha_{1}}'' + e^{-i\chi_{1}t}{\alpha_{3}}'' \Bigr\}\nonumber\\
&& + e^{i\delta}|g_{1}\big\rangle |e_{2}\big\rangle |g_{3}\big\rangle \Bigr\{ {\alpha_{1}}' - e^{-i\chi_{1}t}{\alpha_{3}}' + {\alpha_{1}}'' - e^{-i\chi_{1}t}{\alpha_{3}}'' \Bigr\}\nonumber\\
&& -e^{i\delta}|f_{1}\big\rangle |g_{2}\big\rangle |g_{3}\big\rangle \Bigr\{ {\alpha_{2}}' + e^{-i\chi_{1}t}{\alpha_{4}}' + {\alpha_{2}}'' + e^{-i\chi_{1}t}{\alpha_{4}}'' \Bigr\}\nonumber\\
&& -e^{2i\delta}|f_{1}\big\rangle |e_{2}\big\rangle |g_{3}\big\rangle \Bigr\{ {\alpha_{2}}' - e^{-i\chi_{1}t}{\alpha_{4}}' + {\alpha_{2}}'' - e^{-i\chi_{1}t}{\alpha_{4}}'' \Bigr\}\nonumber\\
&& +e^{i\delta}|g_{1}\big\rangle |g_{2}\big\rangle |e_{3}\big\rangle \Bigr\{ {\alpha_{1}}' + e^{-i\chi_{1}t}{\alpha_{3}}' - {\alpha_{1}}'' - e^{-i\chi_{1}t}{\alpha_{3}}'' \Bigr\}\nonumber\\
&& +e^{2i\delta}|g_{1}\big\rangle |e_{2}\big\rangle |e_{3}\big\rangle \Bigr\{ {\alpha_{1}}' - e^{-i\chi_{1}t}{\alpha_{3}}' - {\alpha_{1}}'' + e^{-i\chi_{1}t}{\alpha_{3}}'' \Bigr\}\nonumber\\
&& -e^{2i\delta}|f_{1}\big\rangle |g_{2}\big\rangle |e_{3}\big\rangle \Bigr\{ {\alpha_{2}}' + e^{-i\chi_{1}t}{\alpha_{4}}' - {\alpha_{2}}'' - e^{-i\chi_{1}t}{\alpha_{4}}'' \Bigr\}\nonumber\\
&& -e^{3i\delta}|f_{1}\big\rangle |e_{2}\big\rangle |e_{3}\big\rangle \Bigr\{ {\alpha_{2}}' - e^{-i\chi_{1}t}{\alpha_{4}}' - {\alpha_{2}}'' + e^{-i\chi_{1}t}{\alpha_{4}}'' \Bigr\}
\Bigg],
\label{ch5finalwave}
\end{eqnarray}
\end{widetext}
which has total $64$ terms, each being the tensor product of five states (three atoms and two fields) with the following coefficients of the entangled fields:
\begin{eqnarray*}
{\alpha_{1}}' & = & \Bigr\{|\alpha_{1}e^{2i\chi_{1}t}\big\rangle|\alpha_{2}e^{2i\chi_{2}t}\big\rangle - |\alpha_{1}e^{i\chi_{1}t}\big\rangle|\alpha_{2}e^{i\chi_{2}t}\big\rangle\Bigr\}, \nonumber\\
{\alpha_{2}}' & = & \Bigr\{|\alpha_{1}e^{2i\chi_{1}t}\big\rangle|\alpha_{2}e^{i\chi_{2}t}\big\rangle + |\alpha_{1}e^{i\chi_{1}t}\big\rangle|\alpha_{2}e^{2i\chi_{2}t}\big\rangle\Bigr\},\nonumber\\
{\alpha_{3}}' & = & \Bigr\{|\alpha_{1}\big\rangle|\alpha_{2}e^{i\chi_{2}t}\big\rangle - |\alpha_{1}e^{-i\chi_{1}t}\big\rangle|\alpha_{2}e^{2i\chi_{2}t}\big\rangle \Bigr\},\nonumber\\
{\alpha_{4}}' & = & \Bigr\{|\alpha_{1}\big\rangle|\alpha_{2}e^{i\chi_{2}t}\big\rangle + |\alpha_{1}e^{-i\chi_{1}t}\big\rangle|\alpha_{2}e^{2i\chi_{2}t}\big\rangle\Bigr\},\nonumber\\
{\alpha_{1}}'' & = & e^{-i\chi_{2}t} \Bigr\{|\alpha_{1}e^{2i\chi_{1}t}\big\rangle|\alpha_{2}e^{-i\chi_{2}t}\big\rangle - |\alpha_{1}e^{i\chi_{1}t}\big\rangle|\alpha_{2}\big\rangle\Bigr\},\nonumber\\
{\alpha_{2}}'' & = & e^{-i\chi_{2}t} \Bigr\{|\alpha_{1}e^{2i\chi_{1}t}\big\rangle|\alpha_{2}e^{-i\chi_{2}t}\big\rangle + |\alpha_{1}e^{i\chi_{1}t}\big\rangle|\alpha_{2}\big\rangle\Bigr\},\nonumber\\
{\alpha_{3}}'' & = & e^{-i\chi_{2}t} \Bigr\{|\alpha_{1}\big\rangle|\alpha_{2}e^{-i\chi_{2}t}\big\rangle - |\alpha_{1}e^{-i\chi_{1}t}\big\rangle|\alpha_{2}\big\rangle\Bigr\},\nonumber\\
{\alpha_{4}}'' & = & e^{-i\chi_{2}t} \Bigr\{|\alpha_{1}\big\rangle|\alpha_{2}e^{-i\chi_{2}t}\big\rangle + |\alpha_{1}e^{-i\chi_{1}t}\big\rangle|\alpha_{2}\big\rangle\Bigr\}.
\end{eqnarray*}
\\
Table \ref{atomfieldstates} lists the atomic state and field state separately, which will offer the entangled superposed field states based on a chosen atomic measurement from eight combinations. From these states, we obtain the entangled states of maximally nonlocal Schr\"{o}dinger cats of both the cavities with $\chi_1 t=\chi_1 t=\pi/2$ from Table \ref{atomfieldstates}. We use these specific case and obtain eight entangled states of horizontal and vertical Schr\"{o}dinger cats:
\begin{eqnarray}
|\psi_{1}\big\rangle & = (-i)\frac{N_{1}}{8}\bigg[\Bigr(|\alpha_{1}\big\rangle + i|-\alpha_{1}\big\rangle\Bigr)\Bigr(|i\alpha_{2}\big\rangle - i|-i\alpha_{2}\big\rangle\Bigr)\nonumber\\
& - \Bigr(|i\alpha_{1}\big\rangle - i|-i\alpha_{1}\big\rangle\Bigr)\Bigr(|\alpha_{2}\big\rangle + i|-\alpha_{2}\big\rangle\Bigr)\bigg],
\label{$g1g2g3$}
\end{eqnarray}
\begin{eqnarray}
|\psi_{2}\big\rangle & = (i)\frac{N_{2}}{8}\bigg[\Bigr(|\alpha_{1}\big\rangle - i|-\alpha_{1}\big\rangle\Bigr)\Bigr(|i\alpha_{2}\big\rangle - i|-i\alpha_{2}\big\rangle\Bigr)\nonumber\\
& + \Bigr(|i\alpha_{1}\big\rangle + i|-i\alpha_{1}\big\rangle\Bigr)\Bigr(|\alpha_{2}\big\rangle + i|-\alpha_{2}\big\rangle\Bigr)\bigg],
\label{$g1e2g3$}
\end{eqnarray}
\begin{eqnarray}
|\psi_{3}\big\rangle & = (i)\frac{N_{3}}{8}\bigg[\Bigr(|\alpha_{1}\big\rangle + i|-\alpha_{1}\big\rangle\Bigr)\Bigr(|i\alpha_{2}\big\rangle - i|-i\alpha_{2}\big\rangle\Bigr)\nonumber\\
& + \Bigr(|i\alpha_{1}\big\rangle - i|-i\alpha_{1}\big\rangle\Bigr)\Bigr(|\alpha_{2}\big\rangle + i|-\alpha_{2}\big\rangle\Bigr)\bigg],
\label{$f1g2g3$}
\end{eqnarray}
\begin{eqnarray}
|\psi_{4}\big\rangle & = (-i)\frac{N_{4}}{8}\bigg[\Bigr(|\alpha_{1}\big\rangle - i|-\alpha_{1}\big\rangle\Bigr)\Bigr(|i\alpha_{2}\big\rangle - i|-i\alpha_{2}\big\rangle\Bigr)\nonumber\\
& - \Bigr(|i\alpha_{1}\big\rangle + i|-i\alpha_{1}\big\rangle\Bigr)\Bigr(|\alpha_{2}\big\rangle + i|-\alpha_{2}\big\rangle\Bigr)\bigg],
\label{$f1e2g3$}
\end{eqnarray}
\begin{eqnarray}
|\psi_{5}\big\rangle & = (-i)\frac{N_{5}}{8}\bigg[\Bigr(|\alpha_{1}\big\rangle + i|-\alpha_{1}\big\rangle\Bigr)\Bigr(|i\alpha_{2}\big\rangle + i|-i\alpha_{2}\big\rangle\Bigr)\nonumber\\
& + \Bigr(|i\alpha_{1}\big\rangle - i|-i\alpha_{1}\big\rangle\Bigr)\Bigr(|\alpha_{2}\big\rangle - i|-\alpha_{2}\big\rangle\Bigr)\bigg],
\label{$g1g2e3$}
\end{eqnarray}
\begin{eqnarray}
|\psi_{6}\big\rangle & = (i)\frac{N_{6}}{8}\bigg[\Bigr(|\alpha_{1}\big\rangle - i|-\alpha_{1}\big\rangle\Bigr)\Bigr(|i\alpha_{2}\big\rangle + i|-i\alpha_{2}\big\rangle\Bigr)\nonumber\\
& - \Bigr(|i\alpha_{1}\big\rangle + i|-i\alpha_{1}\big\rangle\Bigr)\Bigr(|\alpha_{2}\big\rangle - i|-\alpha_{2}\big\rangle\Bigr)\bigg],
\label{$g1e2e3$}
\end{eqnarray}
\begin{eqnarray}
|\psi_{7}\big\rangle & = (i)\frac{N_{7}}{8}\bigg[\Bigr(|\alpha_{1}\big\rangle + i|-\alpha_{1}\big\rangle\Bigr)\Bigr(|i\alpha_{2}\big\rangle + i|-i\alpha_{2}\big\rangle\Bigr)\nonumber\\
& - \Bigr(|i\alpha_{1}\big\rangle - i|-i\alpha_{1}\big\rangle\Bigr)\Bigr(|\alpha_{2}\big\rangle - i|-\alpha_{2}\big\rangle\Bigr)\bigg],
\label{$f1g2e3$}
\end{eqnarray}
\begin{eqnarray}
|\psi_{8}\big\rangle & = (-i)\frac{N_{8}}{8}\bigg[\Bigr(|\alpha_{1}\big\rangle - i|-\alpha_{1}\big\rangle\Bigr)\Bigr(|i\alpha_{2}\big\rangle + i|-i\alpha_{2}\big\rangle\Bigr)\nonumber\\
& + \Bigr(|i\alpha_{1}\big\rangle + i|-i\alpha_{1}\big\rangle\Bigr)\Bigr(|\alpha_{2}\big\rangle - i|-\alpha_{2}\big\rangle\Bigr)\bigg].
\label{$f1e2e3$}
\end{eqnarray}
All of these states involve the entanglement between a position cat and a momentum cat.
\\
\\
\section{Appendix-II}

The Wigner function for the quantum state as given in Eq. (\ref{ch5gencatstate}) is written as
\begin{widetext}
\begin{eqnarray}
W(x_{1},p_{1};x_{2},p_{2}) =  \frac{e^{-\big(|\alpha_{1}|^{2}+|\alpha_{2}|^{2}+x_{1}^{2}+x_{2}^{2}+p_{1}^{2}+
p_{2}^{2}\big)}}{64{\pi}^2} \times\bigg[W_{D1}+W_{D2}+W_{OD1}+W_{OD2}\bigg].\nonumber
\label{ch5totalwigner}
\end{eqnarray}
where $W_{D1}$, $W_{D2}$ are diagonal terms and $W_{OD1}$, $W_{OD2}$ are off-diagonal components, signifying interferences. Their complete expressions are derived to give the terms as follows.
\begin{eqnarray}
W_{D1}  =&& \frac{|A|^{2}}{(1+e^{-2|\alpha_{1}|^{2}}\cos{\xi_{1}})
(1+e^{-2|\alpha_{2}|^{2}}\cos{\zeta_{2}})}\nonumber\\
&& \bigg[e^{-|\alpha_{1}|^{2}}\cos\big\{\sqrt{2}ix_{1}\big(\alpha_{1}e^{i\theta_{1}}+
\alpha^{\ast}_{1}e^{-i\theta_{1}}\big)+\sqrt{2}p_{1}\big(\alpha_{1}e^{i\theta_{1}}-
\alpha^{\ast}_{1}e^{-i\theta_{1}}\big)\big\}\nonumber\\
&& +e^{|\alpha_{1}|^{2}}\cos\big\{\xi_{1}+\sqrt{2}ix_{1}\big(\alpha_{1}e^{i\theta_{1}}-
\alpha^{\ast}_{1}e^{-i\theta_{1}}\big)+\sqrt{2}p_{1}\big(\alpha_{1}e^{i\theta_{1}}+
\alpha^{\ast}_{1}e^{-i\theta_{1}}\big)\big\}\bigg]\nonumber\\
&& \bigg[e^{-|\alpha_{2}|^{2}}\cos\big\{\sqrt{2}ix_{2}\big(\alpha_{2}e^{i\phi_{2}}+
\alpha^{\ast}_{2}e^{-i\phi_{2}}\big)+\sqrt{2}p_{2}\big(\alpha_{2}e^{i\phi_{2}}-
\alpha^{\ast}_{2}e^{-i\phi_{2}}\big)\big\}\nonumber\\
&& +e^{|\alpha_{2}|^{2}}\cos\big\{\zeta_{2}+\sqrt{2}ix_{2}\big(\alpha_{2}e^{i\phi_{2}}-
\alpha^{\ast}_{2}e^{-i\phi_{2}}\big)+\sqrt{2}p_{2}\big(\alpha_{2}e^{i\phi_{2}}+
\alpha^{\ast}_{2}e^{-i\phi_{2}}\big)\big\}\bigg],\nonumber
\label{ch5D1}
\end{eqnarray}
\begin{eqnarray}
W_{D2} = &&
\frac{|B|^{2}}{(1+e^{-2|\alpha_{1}|^{2}}\cos{\zeta_{1}})
(1+e^{-2|\alpha_{2}|^{2}}\cos{\xi_{2}})}\nonumber\\
&& \bigg[e^{-|\alpha_{1}|^{2}}\cos\big\{\sqrt{2}ix_{1}\big(\alpha_{1}e^{i\phi_{1}}+
\alpha^{\ast}_{1}e^{-i\phi_{1}}\big)+\sqrt{2}p_{1}\big(\alpha_{1}e^{i\phi_{1}}-
\alpha^{\ast}_{1}e^{-i\phi_{1}}\big)\big\}\nonumber\\
&& +e^{|\alpha_{1}|^{2}}\cos\big\{\zeta_{1}+\sqrt{2}ix_{1}\big(\alpha_{1}e^{i\phi_{1}}-
\alpha^{\ast}_{1}e^{-i\phi_{1}}\big)+\sqrt{2}p_{1}\big(\alpha_{1}e^{i\phi_{1}}+
\alpha^{\ast}_{1}e^{-i\phi_{1}}\big)\big\}\bigg]\nonumber\\
&& \bigg[e^{-|\alpha_{2}|^{2}}\cos\big\{\sqrt{2}ix_{2}\big(\alpha_{2}e^{i\theta_{2}}+
\alpha^{\ast}_{2}e^{-i\theta_{2}}\big)+\sqrt{2}p_{2}\big(\alpha_{2}e^{i\theta_{2}}-
\alpha^{\ast}_{2}e^{-i\theta_{2}}\big)\big\}\nonumber\\
&& +e^{|\alpha_{2}|^{2}}\cos\big\{\xi_{2}+\sqrt{2}ix_{2}\big(\alpha_{2}e^{i\theta_{2}}-
\alpha^{\ast}_{2}e^{-i\theta_{2}}\big)+\sqrt{2}p_{2}\big(\alpha_{2}e^{i\theta_{2}}+
\alpha^{\ast}_{2}e^{-i\theta_{2}}\big)\big\}\bigg],\nonumber
\label{ch5D2}
\end{eqnarray}
\begin{eqnarray}
W_{OD1}  = && \frac{(A\times B^{\ast})exp[\frac{i((\xi_{1}-\xi_{2})-(\zeta_{1}-\zeta_{2}))}{2}]}
{\sqrt{(1+e^{-2|\alpha_{1}|^{2}}\cos{\xi_{1}})(1+e^{-2|\alpha_{2}|^{2}}\cos{\zeta_{2}})
(1+e^{-2|\alpha_{1}|^{2}}\cos{\zeta_{1}})(1+e^{-2|\alpha_{2}|^{2}}\cos{\xi_{2}})}}\nonumber\\
&& \bigg[e^{-|\alpha_{1}|^{2}e^{i(\theta_{1}-\phi_{1})}}\cos\big\{\frac{(\zeta_{1}-
\xi_{1})}{2}-\sqrt{2}ix_{1}\big(\alpha_{1}e^{i\theta_{1}}+
\alpha^{\ast}_{1}e^{-i\phi_{1}}\big)-\sqrt{2}p_{1}\big(\alpha_{1}e^{i\theta_{1}}-
\alpha^{\ast}_{1}e^{-i\phi_{1}}\big)\big\}\nonumber\\
&& + e^{|\alpha_{1}|^{2}e^{i(\theta_{1}-\phi_{1})}}\cos\big\{\frac{(\zeta_{1}+
\xi_{1})}{2}+\sqrt{2}ix_{1}\big(\alpha_{1}e^{i\theta_{1}}-
\alpha^{\ast}_{1}e^{-i\phi_{1}}\big)+\sqrt{2}p_{1}\big(\alpha_{1}e^{i\theta_{1}}+
\alpha^{\ast}_{1}e^{-i\phi_{1}}\big)\big\}\bigg]\nonumber\\
&& \bigg[e^{-|\alpha_{2}|^{2}e^{i(\phi_{2}-\theta_{2})}}\cos\big\{\frac{(\zeta_{2}-
\xi_{2})}{2}+\sqrt{2}ix_{2}\big(\alpha_{2}e^{i\phi_{2}}+
\alpha^{\ast}_{2}e^{-i\theta_{2}}\big)+\sqrt{2}p_{2}\big(\alpha_{2}e^{i\phi_{2}}-
\alpha^{\ast}_{2}e^{-i\theta_{2}}\big)\big\}\nonumber\\
&& + e^{|\alpha_{2}|^{2}e^{i(\phi_{2}-\theta_{2})}}\cos\big\{\frac{(\zeta_{2}+
\xi_{2})}{2}+\sqrt{2}ix_{2}\big(\alpha_{2}e^{i\phi_{2}}-
\alpha^{\ast}_{2}e^{-i\theta_{2}}\big)+\sqrt{2}p_{2}\big(\alpha_{2}e^{i\phi_{2}}+
\alpha^{\ast}_{2}e^{-i\theta_{2}}\big)\big\}\bigg],\nonumber
\label{ch5OD1}
\end{eqnarray}
and
\begin{eqnarray}
W_{OD2} = && \frac{(A^{\ast}\times B)exp[\frac{i((\zeta_{1}-\zeta_{2})-(\xi_{1}-\xi_{2}))}{2}]}
{\sqrt{(1+e^{-2|\alpha_{1}|^{2}}\cos{\xi_{1}})(1+e^{-2|\alpha_{2}|^{2}}\cos{\zeta_{2}})
(1+e^{-2|\alpha_{1}|^{2}}\cos{\zeta_{1}})(1+e^{-2|\alpha_{2}|^{2}}\cos{\xi_{2}})}}\nonumber\\
&& \bigg[e^{-|\alpha_{1}|^{2}e^{i(\phi_{1}-\theta_{1})}}\cos\big\{\frac{(\xi_{1}-
\zeta_{1})}{2}-\sqrt{2}ix_{1}\big(\alpha_{1}e^{i\phi_{1}}+
\alpha^{\ast}_{1}e^{-i\theta_{1}}\big)-\sqrt{2}p_{1}\big(\alpha_{1}e^{i\phi_{1}}-
\alpha^{\ast}_{1}e^{-i\theta_{1}}\big)\big\}\nonumber\\
&& + e^{|\alpha_{1}|^{2}e^{i(\phi_{1}-\theta_{1})}}\cos\big\{\frac{(\xi_{1}+
\zeta_{1})}{2}+\sqrt{2}ix_{1}\big(\alpha_{1}e^{i\phi_{1}}-
\alpha^{\ast}_{1}e^{-i\theta_{1}}\big)+\sqrt{2}p_{1}\big(\alpha_{1}e^{i\phi_{1}}+
\alpha^{\ast}_{1}e^{-i\theta_{1}}\big)\big\}\bigg]\nonumber\\
&& \bigg[e^{-|\alpha_{2}|^{2}e^{i(\theta_{2}-\phi_{2})}}\cos\big\{\frac{(\xi_{2}-
\zeta_{2})}{2}+\sqrt{2}ix_{2}\big(\alpha_{2}e^{i\theta_{2}}+
\alpha^{\ast}_{2}e^{-i\phi_{2}}\big)+\sqrt{2}p_{2}\big(\alpha_{2}e^{i\theta_{2}}-
\alpha^{\ast}_{2}e^{-i\phi_{2}}\big)\big\}\nonumber\\
&& + e^{|\alpha_{2}|^{2}e^{i(\theta_{2}-\phi_{2})}}\cos\big\{\frac{(\xi_{2}+
\zeta_{2})}{2}+\sqrt{2}ix_{2}\big(\alpha_{2}e^{i\theta_{2}}-
\alpha^{\ast}_{2}e^{-i\phi_{2}}\big)+\sqrt{2}p_{2}\big(\alpha_{2}e^{i\theta_{2}}+
\alpha^{\ast}_{2}e^{-i\phi_{2}}\big)\big\}\bigg].\nonumber
\label{ch5OD2}
\end{eqnarray}
\end{widetext}


\begin{thebibliography}{99}
\bibitem{einstein1935can} A. Einstein, B. Podolsky, N. Rosen, Physical Review {\bf 47}  777 (1935).
\bibitem{clauser1978bell} J. F. Clauser, A. Shimony, Reports on Progress in Physics {\bf 41}  1881 (1978).
\bibitem{aspect1982experimental} A. Aspect, P. Grangier, G. Roger, Physical Review Letters {\bf 49}  91 (1982).
\bibitem{mair2001entanglement} A. Mair, A. Vaziri, G. Weihs, A. Zeilinger, Nature {\bf 412}  313–316 (2001).
\bibitem{greenberger1990bell} D. M. Greenberger, M. A. Horne, A. Shimony, A. Zeilinger, American Journal of Physics {\bf 58}  1131–1143 (1990).
\bibitem{erhard2020advances} M. Erhard, M. Krenn, A. Zeilinger, Nature Reviews Physics {\bf 2}  365–381 (2020).
\bibitem{shukla2019quantum} N. Shukla, N. Akhtar, B. C. Sanders, Physical Review A {\bf 99}  063813 (2019).
\bibitem{roy2009sub} U. Roy, S. Ghosh, P. K. Panigrahi, D. Vitali, Physical Review A {\bf 80} 052115 (2009).
\bibitem{buvzek1992superpositions} V. Buvzek, A. Vidiella-Barranco, P. L. Knight, Physical Review A {\bf 45}  6570 (1992).
\bibitem{sanders2012review} B. C. Sanders, Journal of Physics A: Mathematical and Theoretical {\bf 45}  244002 (2012).
\bibitem{sun1992generation} J. Sun, J. Wang, C. Wang, Physical Review A {\bf 46}  1700 (1992).
\bibitem{yurke1986generating} B. Yurke, D. Stoler, Physical Review Letters {\bf 57}  13 (1986).
\bibitem{agarwal1992new} G. Agarwal, R. Simon, Optics Communications {\bf 92} 105–107 (1992).
\bibitem{solomon1994characteristic} A. I. Solomon, Physics Letters A {\bf 196} 29–34 (1994).
\bibitem{wigner1997quantum} E. P. Wigner, in: Part I: Physical Chemistry. Part II: Solid State Physics, Springer, pp. 110–120 (1997).
\bibitem{hillery1984distribution} M. Hillery, R. F. O’Connell, M. O. Scully, E. P. Wigner, Physics Reports {\bf 106} 121–167 (1984).
\bibitem{kenfack2004negativity}  A. Kenfack, K. Zyczkowski, Journal of Optics B: Quantum and Semiclassical Optics {\bf 6} 396 (2004).
\bibitem{van2001entangled} S. J. van Enk, O. Hirota, Physical Review A {\bf 64} 022313 (2001).
\bibitem{horoshko2019quantum} D. Horoshko, G. Patera, M. Kolobov, Optics Communications {\bf 447} 67–73 (2019).
\bibitem{ghosh2006mesoscopic} S. Ghosh, A. Chiruvelli, J. Banerji, P. Panigrahi, Physical Review A {\bf 73} 013411 (2006).
\bibitem{ghosh2009sub} S. Ghosh, U. Roy, C. Genes, D. Vitali, Physical Review A {\bf 79} 052104 (2009).
\bibitem{ghosh2014enhanced} S. Ghosh, U. Roy, Physical Review A {\bf 90} 022113 (2014).
\bibitem{ghosh2019sub} S. Ghosh, J. Bera, P. K. Panigrahi, U. Roy, International Journal of Quantum Information {\bf 17} 1950019 (2019).
\bibitem{akhtar2021sub} N. Akhtar, B. C. Sanders, C. Navarrete-Benlloch, Physical Review A {\bf 103} 053711 (2021).
\bibitem{agarwal2022quantifying} G. Agarwal, L. Davidovich, Physical Review Research {\bf 4} L012014 (2022).
\bibitem{bera2020matter} J. Bera, S. Ghosh, L. Salasnich, U. Roy, Physical Review A {\bf 102} 063323 (2020).
\bibitem{bera2022quantum} J. Bera, B. Halder, S. Ghosh, R.-K. Lee, U. Roy, Physics Letters A {\bf 453} 128484 (2022).
\bibitem{joo2011quantum}  J. Joo, W. J. Munro, T. P. Spiller, Physical Review Letters {\bf 107} 083601 (2011).
\bibitem{cappellaro2005entanglement} P. Cappellaro, J. Emerson, N. Boulant, C. Ramanathan, S. Lloyd, D. G. Cory, Physical Review Letters {\bf 94} 020502 (2005).
\bibitem{demkowicz2014using} R. Demkowicz-Dobrza´nski, L. Maccone, Physical Review Letters {\bf 113} 250801 (2014).
\bibitem{huang2016usefulness} Z. Huang, C. Macchiavello, L. Maccone, Physical Review A {\bf 94} 012101 (2016).
\bibitem{cleve1997substituting} R. Cleve, H. Buhrman, Physical Review A {\bf 56} 1201 (1997).
\bibitem{bostrom2002deterministic} K. Bostr¨om, T. Felbinger, Physical Review Letters {\bf 89} 187902 (2002).
\bibitem{hastings2009superadditivity} M. B. Hastings, Nature Physics {\bf 5} 255–257 (2009).

\bibitem{Otta} K. N. Wilkinson, P. Papanastasiou, C. Ottaviani, T. Gehring, S. Pirandola, Physical Review Research {\bf 2} 033424 (2020).
\bibitem{tittel2000quantum} W. Tittel, J. Brendel, H. Zbinden, N. Gisin, Physical Review Letters {\bf 84} 4737 (2000).
\bibitem{yin2020entanglement} J. Yin, Y.-H. Li, S.-K. Liao, M. Yang, Y. Cao, L. Zhang, J.-G. Ren, W.-Q. Cai, W.-Y. Liu, S.-L. Li, et al., Nature {\bf 582} 501–505 (2020).
\bibitem{jennewein2000quantum} T. Jennewein, C. Simon, G. Weihs, H. Weinfurter, A. Zeilinger, Physical Review Letters {\bf 84} 4729 (2000).
\bibitem{zidan2020novel} M. Zidan, Modern Physics Letters B {\bf 34} 2050401 (2020).
\bibitem{li2014triple} Q. Li, W. H. Chan, C. Wu, Z. Wen, Physical Review A {\bf 89} 040302 (2014).
\bibitem{ding2007review} S. Ding, Z. Jin, Chinese Science Bulletin {\bf 52} 2161–2166 (2007).
\bibitem{ottaviani2010}C. Ottaviani and D. Vitali,
Phys. Rev. A \textbf{82}, 012319 (2010).
\bibitem{mirrahimi2023}J. Guillaud, J. Cohen, and M. Mirrahimi, SciPost Phys. Lect. Notes 72 (2023)
\bibitem{lo2001concentrating} H.-K. Lo, S. Popescu, Physical Review A {\bf 63} 022301 (2001).
\bibitem{dur2000three} W. Dur, G. Vidal, J. I. Cirac, Physical Review A {\bf 62} 062314 (2000).
\bibitem{acin2001classification} A. Acın, D. Bruß, M. Lewenstein, A. Sanpera, Physical Review Letters {\bf 87} 040401 (2001).
\bibitem{nielsen2006cluster} M. A. Nielsen, Reports on Mathematical Physics {\bf 57} 147–161 (2006).
\bibitem{walther2005experimental} P. Walther, M. Aspelmeyer, K. J. Resch, A. Zeilinger, Physical Review Letters {\bf 95} 020403 (2005) .
\bibitem{briegel2001persistent} H. J. Briegel, R. Raussendorf, Physical Review Letters {\bf 86} 910 (2001).
\bibitem{haroche1989cavity} S. Haroche, D. Kleppner, Phys. Today {\bf 42} 24–30 (1989).
\bibitem{wang2019turning} D. Wang, H. Kelkar, D. Martin-Cano, D. Rattenbacher, A. Shkarin, T. Utikal, S. G¨otzinger, V. Sandoghdar, Nature Physics {\bf 15} 483–489 (2019).
\bibitem{davidovich1996mesoscopic} L. Davidovich, M. Brune, J. Raimond, S. Haroche, Physical Review A {\bf 53} 1295 (1996).
\bibitem{brune1996observing} M. Brune, E. Hagley, J. Dreyer, X. Maitre, A. Maali, C. Wunderlich, J. Raimond, S. Haroche, Physical Review Letters {\bf 77} 4887 (1996).
\bibitem{raimond1997reversible} J. Raimond, M. Brune, S. Haroche, Physical Review Letters {\bf 79} 1964 (1997).
\bibitem{raimond2001manipulating} J.-M. Raimond, M. Brune, S. Haroche, Reviews of Modern Physics {\bf 73} 565 (2001).
\bibitem{agarwal1997atomic} G. S. Agarwal, R. Puri, R. Singh, Physical Review A {\bf 56} 2249 (1997).
\bibitem{gerry1996generation} C. C. Gerry, Physical Review A {\bf 53} 3818 (1996).
\bibitem{montina2002bistability} A. Montina, F. Arecchi, Physical Review A {\bf 66} 013605 (2002) .
\bibitem{solano2003generalized} E. Solano, G. S. Agarwal, H. Walther, Optics and Spectroscopy {\bf 94} 805–807 (2003).
\bibitem{chai1992two} C.-L. Chai, Physical Review A {\bf 46} 7187 (1992).
\bibitem{mogilevtsev1996generation} D. Mogilevtsev, S. Y. Kilin, Optics Communications {\bf 132} 452–456 (1996).
\bibitem{vitali2000generating} D. Vitali, M. Fortunat, P. Tombesi, F. De Martini, Fortschritte der Physik: Progress of Physics {\bf 48} 437–446 (2000).
\bibitem{kanari2022two} L. A. Kanari-Naish, J. Clarke, S. Qvarfort, M. R. Vanner, Quantum Science and Technology {\bf 7} 035012 (2022).
\bibitem{huang2020generation} J. Huang, Y.-H. Liu, J.-F. Huang, J.-Q. Liao, Physical Review A {\bf 101} 043841 (2020).
\bibitem{zurek2001sub} W. H. Zurek, Nature {\bf 412} 712–717 (2001).
\bibitem{jacquod2002decay} P. Jacquod, I. Adagideli, C. W. Beenakker, Physical Review Letters {\bf 89} 154103  (2002).
\bibitem{wisniacki2003short} D. A. Wisniacki, Physical Review E {\bf 67} 016205 (2003).
\bibitem{pathak2005generation} P. Pathak, G. Agarwal, Physical Review A {\bf 71} 043823 (2005).
\bibitem{toscano2006sub} F. Toscano, D. A. Dalvit, L. Davidovich, W. H. Zurek, Physical Review A {\bf 73} 023803 (2006).
\bibitem{dalvit2006quantum} D. Dalvit, R. de Matos Filho, F. Toscano, New Journal of Physics {\bf 8} 276 (2006).
\bibitem{agarwal2004mesoscopic} G. Agarwal, P. Pathak, Physical Review A {\bf 70} 053813 (2004).
\bibitem{bhatt2008entanglement} J. R. Bhatt, P. K. Panigrahi, M. Vyas, Physical Review A {\bf 78} 034101 (2008).
\bibitem{stobinska2008wigner} M. Stobinska, G. Milburn, K. W´odkiewicz, Physical Review A {\bf 78} 013810 (2008).
\bibitem{banerji2007exploring} J. Banerji, Contemporary Physics {\bf 48} 157–171 (2007).
\bibitem{hirota2001entangled} O. Hirota, S. J. Van Enk, K. Nakamura, M. Sohma, K. Kato, arXiv preprint quant-ph/0101096 (2001).
\bibitem{hirota2002entangled} O. Hirota, M. Sasaki, Quantum Communication, Computing, and Measurement {\bf 3} 359–366 (2002).
\bibitem{ahmad2006entangled} M. A. Ahmad, L. Shu-Tian, Chinese Physics Letters {\bf 23} 2964 (2006).
\bibitem{zeng2007nonclassical} R. Zeng, M. A. Ahmad, S. Liu, Optics Communications {\bf 271} 162–168 (2007).
\bibitem{bukhari2011two} S. H. Bukhari, S. N. Khan, M. A. Ahmad, Acta Phys. Pol. B {\bf 42} 2077–2086 (2011).
\bibitem{hirota2017error}  O. Hirota, Quantum Measurements and Quantum Metrology {\bf 4} 70–73 (2017).
\bibitem{kato2015quasi} K. Kato, in: Quantum Communications and Quantum Imaging XIII, volume 9615, SPIE, pp. 65–74
\bibitem{Horodecki}R. Horodecki, P. Horodecki, M. Horodecki, and K. Horodecki, Rev. Mod. Phys. {\bf 81}, 865  (2009).
\bibitem{Mari2009}A. Mari, and J. Eisert, Phys. Rev. Lett. \textbf{109}, 230503 (2012).
\bibitem{Walschaers2023}M. Walschaers, Quantum \textbf{7}, 1038 (2023).

\end{thebibliography}
\end{document}